\documentclass[aps,prl,reprint,superscriptaddress,longbibliography,nofootinbib,floatfix]{revtex4-2}

\usepackage{amsmath,amssymb,bm}
\usepackage{graphicx}
\usepackage{microtype}
\usepackage{xcolor}
\usepackage[colorlinks=true,allcolors=blue]{hyperref}

\newcommand{\Tr}{\operatorname{Tr}}
\newcommand{\Diss}[1]{\mathcal{D}[#1]}

\newcommand{\dd}{\mathrm{d}}
\newcommand{\ee}{\mathrm{e}}
\newcommand{\ketbra}[2]{|#1\rangle\!\langle#2|}
\newcommand{\proj}[1]{|#1\rangle\!\langle#1|}

\begin{document}

\title{Thermodynamics of Speed and Efficiency with Quantum Feedback}

\author{Yue Liu}
\email{yue.liu@yukawa.kyoto-u.ac.jp}
\affiliation{Center for Gravitational Physics and Quantum Information, Yukawa Institute for Theoretical Physics, Kyoto University, Kitashirakawa Oiwakecho, Sakyo-ku, Kyoto 606-8502, Japan}

\author{Ryotaro Honma}
\affiliation{Center for Gravitational Physics and Quantum Information, Yukawa Institute for Theoretical Physics, Kyoto University, Kitashirakawa Oiwakecho, Sakyo-ku, Kyoto 606-8502, Japan}

\author{Ruicheng Bao}
\affiliation{Department of Physics, Graduate School of Science, The University of Tokyo, Hongo, Bunkyo-ku, Tokyo 113-0033, Japan}

\author{Tan Van Vu}
\email{tan.vu@yukawa.kyoto-u.ac.jp}
\affiliation{Center for Gravitational Physics and Quantum Information, Yukawa Institute for Theoretical Physics, Kyoto University, Kitashirakawa Oiwakecho, Sakyo-ku, Kyoto 606-8502, Japan}

\begin{abstract}
Rapid thermodynamic operations generally entail irreversible dissipation, giving rise to trade-offs between operating speed and thermodynamic cost. 
Yet, how quantum feedback fundamentally reshapes these trade-offs remains elusive.
Here, we derive a finite-time current--dissipation relation for continuously monitored open quantum dynamics with jump feedback. 
From this fundamental relation, we establish a quantum speed limit and a power-efficiency trade-off for stationary information engines, identifying how feedback affects state transformations and the performance of engines through both informational entropy and direct dynamical contributions. 
Our results provide a quantitative foundation for fast, low-dissipation quantum thermodynamics in the presence of feedback control, exemplified by attaining the generalized Landauer bound in finite-time information erasure and approaching unit information efficiency at finite power in the large-system limit of a collective engine.
\end{abstract}

\maketitle

\textit{Introduction.---}Achieving rapid operations with less irreversible losses is a central challenge in thermodynamics~\cite{Seifert2012}. 
The second law sets fundamental limits on thermodynamic tasks, such as the Carnot efficiency for heat engines~\cite{Callen1985} and the Landauer bound for information erasure~\cite{Landauer1961}.
Nevertheless, these benchmarks are strictly attainable only in the quasistatic limit, as finite-time operation inevitably incurs irreversible dissipation. 
Practical heat engines operate well below the Carnot efficiency~\cite{Curzon1975}, while heat dissipation in information processing exceeds the Landauer bound by many orders of magnitude~\cite{Wolpert2024}. 
%The second law alone does not quantify how these losses depend on the operating speed. 
Over the past few decades, advances in stochastic thermodynamics~\cite{Peliti2021,Shiraishi2023} and quantum thermodynamics~\cite{Binder2018,Campbell2026} have established general thermodynamic bounds that tighten the second law and quantify the cost of finite-time operations. 
These include thermodynamic speed limits~\cite{Shiraishi2018,Funo2019,Tan2021,Dechant2022,Tan2023Topology,Tan2023}, thermodynamic and kinetic uncertainty relations~\cite{Barato2015,Gingrich2016,Pietzonka2018,Terlizzi2019,Hasegawa20191,Hasegawa2020,Koyuk2020,Tan20222,Tan2023,Prech2025,Tan2025,Tan2026Precision}, and power-efficiency trade-offs~\cite{Whitney2014,Shiraishi2016,Pietzonka2018,ShiraishiSaito2019,Brandner2020,Tajima2021,Funo2025}, which quantitatively constrain finite-time information erasure~\cite{Proesmans2020L,Miller2020,Zhen2021,Tan20221,Lee2022,Rolandi2023,Liu20253,Tan2025TimeCostError} and energy conversion~\cite{Aamir2025,Liang2025,Hou2025}, as well as the precision of clocks~\cite{Meier2025,Prech2025Clock,Wadhia2025}.

\suppressfloats[t]
\begin{figure}[t]
  \centering
  \includegraphics[width=\columnwidth]{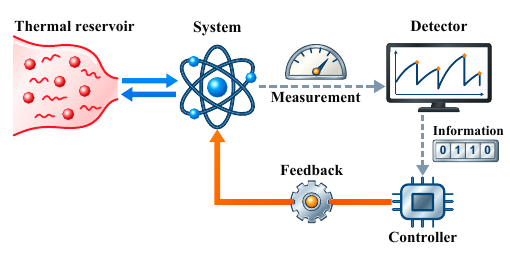}
  \caption{A quantum system exchanges energy with a thermal reservoir. Detected quantum jumps provide information to a controller, which applies the corresponding feedback operation to the system.}
  \label{fig:quantum-feedback-setup}
\end{figure}

The idea of measurement and feedback dates back to Maxwell's demon and the Szilard engine~\cite{Parrondo2015,deOliveiraJunior2025}, where acquired information is harnessed to perform thermodynamic tasks such as work extraction. 
Such processes initially seemed to challenge the second law of thermodynamics, a paradox resolved by incorporating information into the entropy balance~\cite{Sagawa2008,Sagawa2010,Sagawa2012Fluctuation,Yada2022}. 
Theoretical and experimental studies have explored the role of information in thermodynamic processes for specific models~\cite{Saha2021,Fadler2023,Zhou2024,AnnbyAndersson2024,duBuisson2024,Aggarwal2025,Archambault2025,Liu2025Feedback,Kiran2025,Tohme2025,PatronCastro2026,Dago2026}, together with thermodynamic trade-offs~\cite{Nagase2024,Kamijima2025,Kamijima2025Langevin,Nagase2025,Kirchberg2025}. 
The precision of currents can be enhanced by using information and feedback to suppress fluctuations, as demonstrated by thermodynamic uncertainty relations~\cite{Potts2019,Hasegawa2023,Kumasaki2026,HonmaVanVu2026,Tojo2026}. 
These advances raise the question of how quantum feedback modifies finite-time thermodynamic bounds through both information contributions and dynamical effects.

%Addressing this question holds the key to assessing how information can be harnessed to optimize the trade-off between operating speed and dissipation in memory erasure and heat engines.

In this Letter, we establish a current--dissipation relation to clarify the role of Markovian feedback in finite-time quantum thermodynamics. 
From this relation, we derive a quantum speed limit for state transformations and a power-efficiency trade-off for stationary information heat engines. 
The resulting speed limit demonstrates that feedback governs the speed of state transformations not only through the information contribution, but also via a direct dynamical action of the feedback operation. 
Crucially, this direct contribution enables the generalized Landauer bound to be saturated within a finite erasure time. 
Furthermore, when applied to a collective information engine, our power-efficiency trade-off demonstrates that many-body cooperative effects allow the engine to approach the maximum information efficiency while sustaining a finite output power in the thermodynamic limit.

\textit{Setup.---}We consider a quantum system weakly coupled to thermal reservoirs $\{\nu\}$ at inverse temperatures $\{\beta_\nu\}$.
As illustrated in Fig.~\ref{fig:quantum-feedback-setup}, a detector monitors the quantum jumps induced by the reservoirs and feeds the measurement outcomes to a controller, which instantaneously applies the corresponding feedback operation after each detected jump. 
The resulting dynamics is described by the master equation~\cite{Wiseman1994,Wiseman2010},
\begin{equation}
\begin{aligned}
  \dot\rho_t=&-i[H_t,\rho_t]\\
  &+\sum_{\nu,k}\left(
  \mathcal F_{\nu k}[L_{\nu k}\rho_tL_{\nu k}^{\dagger}]-\frac{1}{2}\{L_{\nu k}^{\dagger}L_{\nu k},\rho_t\}\right),
\end{aligned}
\label{FB_Lindblad}
\end{equation}
where $H_t$ denotes the time-dependent Hamiltonian and $L_{\nu k}$ describes a jump in channel $k$ induced by reservoir $\nu$. 
The jump operators satisfy local detailed balance, $L_{\nu k}=\ee^{\Delta s_{\nu k}/2}L_{\nu k^*}^{\dagger}$, where $k^*$ labels the reverse channel and $\Delta s_{\nu k}=-\Delta s_{\nu k^*}$ is the reservoir entropy change associated with the jump. 
The feedback operation $\mathcal F_{\nu k}$, applied conditioned on the detected jump $(\nu,k)$, is a unital quantum channel ($\mathcal{F}_{\nu k}[\mathbb{I}]=\mathbb{I}$), which includes unitary transformations, nonselective projective measurements, and the identity map. 
Such feedback schemes based on quantum jump detection have been realized experimentally in various architectures, ranging from trapped-ion systems~\cite{Akerman2012} to cavity QED setups~\cite{Smith2002,Reiner2004}.

Feedback uses information obtained from quantum jump detection, which must be taken into account when evaluating entropy production. 
For a process of duration $\tau$, the entropy production including this contribution obeys the generalized second law~\cite{HonmaVanVu2026}
\begin{equation}\label{fb_entropy_production}
  \Sigma_\tau:=\Delta S_{\mathrm{sys}}+\Delta S_{\mathrm{env}}-I=\int_0^\tau \sigma_t\,\dd t\geq0,
\end{equation}
where $\sigma_{t}$ denotes the instantaneous entropy production rate.
Here, $\Delta S_{\mathrm{sys}}=S(\rho_\tau)-S(\rho_0)$ is the change in system entropy, with von Neumann entropy $S(\rho)=-\Tr(\rho\ln\rho)$, and $\Delta S_{\mathrm{env}}=\int_0^\tau \dd t\,\sum_{\nu,k}\Delta s_{\nu k}(t)a_{\nu k}(t)$ is the mean entropy change of the thermal reservoirs, with $a_{\nu k}(t)=\Tr(L_{\nu k}\rho_tL_{\nu k}^{\dagger})$ being the jump rate of channel $(\nu,k)$. 
The information term $I$, which quantifies the amount of information exploited during the feedback step, reads~\cite{HonmaVanVu2026} 
\begin{equation}\label{information}
  I=\int_0^\tau \dd t\,\sum_{\nu,k}a_{\nu k}(t) \left[D(\mathcal F_{\nu k}[\rho'_{\nu k,t}]\Vert\rho_t)-D(\rho'_{\nu k,t}\Vert\rho_t)\right],
\end{equation}
where $D(\cdot\Vert\cdot)$ is the quantum relative entropy and $\rho'_{\nu k,t}=L_{\nu k}\rho_tL_{\nu k}^{\dagger}/a_{\nu k}(t)$ is the post-jump state prior to feedback. Consistent with established treatments~\cite{Strasberg2013PRL,Strasberg2013PRE,Jun2014}, our description focuses on the system and its reservoirs, explicitly accounting for the information contribution $I$.

%\tv{Tan: I suggest explaining the information term further before showing its formulation; this would help the reader better understanding the essence of this quantity.}
%The quantity $\Sigma_{\tau}:=\int_{0}^{\tau}\sigma_{t}\mathrm{d}t$ characterizes the entropy production associated with the system and its reservoirs, accounting for the information used by feedback.

%The change in mutual information is taken after feedback minus before feedback, so information consumption corresponds to $I<0$.
%tv{The transition to this section should be smoother. A reader unfamiliar with this topic may be unclear about what we are doing here and what is its purpose. Starting with some reasoning about the origin of trade-offs in terms of current-dissipation relations may be an option.} 

\textit{Current--dissipation relation.---}The thermodynamic cost of transport sets limits on the performance of heat engines~\cite{Shiraishi2016} and provides insight into the role of coherence and symmetry in transport~\cite{Tajima2021,Funo2025}. 
Here, we establish such a relation for quantum systems with feedback.
We consider the jumps in the instantaneous eigenbasis of the density matrix, $\rho_t=\sum_y p_y(t)\proj{y_t}$ with $p_y(t)>0$. 
In this basis, $w_{\nu k}^{xy}(t):=|\langle x_t|L_{\nu k}|y_t\rangle|^2$ is the transition rate from $|y_t\rangle$ to $|x_t\rangle$ through the jump channel $(\nu,k)$, and $w_{\nu k}^{xy}(t)p_y(t)$ is the corresponding probability flux. 
For real, time-dependent weights satisfying $c_{\nu k^*}^{yx}(t)=-c_{\nu k}^{xy}(t)$, we define the weighted current as
\begin{equation}
  J_t:=\sum_{\nu,k,x,y}c_{\nu k}^{xy}(t)
  w_{\nu k}^{xy}(t)p_y(t).
\label{eq:weighted-current-coefficient}
\end{equation}
The weights $c_{\nu k}^{xy}(t)$ quantify the contribution of each transition, such that $J_t$ represents the corresponding mean net current under feedback control. 
A key result in our study is the following finite-time current--dissipation relation for the feedback dynamics (see Supplemental Material (SM)~\cite{supp} for details)
\begin{equation}
  \left(\int_0^\tau |J_t|\,\dd t\right)^2
  \leq M_\tau\,\Sigma_\tau.
\label{fb_cdb}
\end{equation}
Here, $M_\tau:=\int_0^\tau m_t\,\dd t$, where $m_t$ depends on the forward and reverse probability fluxes and the chosen current weights.
It can be viewed as a weighted dynamical state mobility~\cite{Tan2023}, with its explicit expression provided in the End Matter.

% where the nonnegative rate $m_c(t)$ is determined by the squared weights and the logarithmic means of the forward and reverse probability fluxes, as detailed in Eq.~\eqref{eq:end-matter-general-coefficient} of the End Matter.

This relation strengthens the generalized second law of thermodynamics in Eq.~\eqref{fb_entropy_production} by showing that a nonzero current necessarily entails positive entropy production. It also quantifies how information utilized by feedback reduces the thermodynamic cost of transport. Furthermore, appropriate choices of the weights $c_{\nu k}^{xy}(t)$ yield a diverse family of thermodynamic trade-offs. We discuss two representative applications below. Additional applications, including an information heat pump and a cyclic heat engine, are provided in SM~\cite{supp}.

\textit{Quantum speed limit.---}We first derive a quantum speed limit from the key relation \eqref{fb_cdb}, which bounds the minimum time required for a state transformation. 
To this end, we choose $c_{\nu k}^{xy}=z_x-z_y$, with $z_x=\operatorname{sgn}[p_x(\tau)-p_x(0)]$, and obtain the quantum speed limit~\cite{supp}
\begin{equation}\label{fb_qsl}
  \tau\geq\frac{d_T(\rho_0,\rho_\tau)}{\sqrt{\overline{m}\,\overline\sigma}+\overline b},
\end{equation}
where $\overline{\upsilon}:=(1/\tau)\int_0^\tau\upsilon_t\,\dd t$ is the time average of a scalar quantity $\upsilon_t$. 
Here, $d_T(\rho_0,\rho_\tau):=\sum_y|p_y^\downarrow(\tau)-p_y^\downarrow(0)|$ is the trace-like distance~\cite{Tan2021}, with $p_y^\downarrow(t)$ denoting the eigenvalues of $\rho_t$ arranged in descending order. 
The mean feedback drift is determined by $b_t:=\|\sum_{\nu,k}a_{\nu k}(t)(\mathcal F_{\nu k}[\rho'_{\nu k,t}]-\rho'_{\nu k,t})\|_1$, which quantifies the rate of state change directly induced by feedback, with $\|\cdot\|_1$ denoting the trace norm.
Quantum feedback therefore enters the speed limit through two distinct pathways: the informational contribution to the entropy production rate $\overline\sigma$, and the direct dynamical drift $\overline b$.
%, which enables state transformations to be steered directly by feedback control.

%The trace norm is defined as $\|X\|_1:=\Tr\sqrt{X^{\dagger}X}$. For the chosen weights, the weighted mobility becomes $m_c(t)=\frac12\sum_{\nu,k,x,y}(z_x-z_y)^2\ell_{\nu k}^{xy}(t)$.

Several remarks are in order. 
First, the presence of the feedback drift and the information term suggests that suitably designed feedback can accelerate state transformations, offering potential advantages for finite-time state preparation, such as cooling and information erasure. 
Second, the weighted mobility is bounded from above by the instantaneous dynamical activity, $m_t\leq2a_t$ (see the End Matter), where $a_t:=\sum_{\nu,k}a_{\nu k}(t)$. 
In the absence of feedback ($I=0$ and $\overline b=0$), Eq.~\eqref{fb_qsl} recovers the known quantum and classical speed limits~\cite{Shiraishi2018,Tan2021,Tan2023}. 
%Furthermore, because $m_t\leq2a_t$, our bound is generally tighter than existing speed limits formulated via dynamical activity~\cite{supp}.

\textit{Example I: Finite-time information erasure.---}As an application of the quantum speed limit~\eqref{fb_qsl}, we consider a process of information erasure, where the system is coupled to an environment at inverse temperature $\beta$. 
This fundamental task lowers the system entropy ($\Delta S_{\mathrm{sys}}<0$), e.g., by transforming a maximally mixed state into a pure state, and is generally accompanied by heat dissipation $Q$ into the environment.
% This fundamental thermodynamic task involves resetting the system to a state of lower entropy ($\Delta S_{\mathrm{sys}}<0$), such as transforming a maximally mixed state into a pure state, which is generally accompanied by heat dissipation $Q$ into the environment. 
The generalized second law~\eqref{fb_entropy_production} yields the generalized Landauer bound $\beta Q\geq-\Delta S_{\mathrm{sys}}+I$.
%, which implies that the heat dissipation can be reduced by the information contribution $I$. 
The quantum speed limit~\eqref{fb_qsl} further strengthens this bound by quantifying the additional cost of finite-time operation and dynamical feedback:
\begin{equation}\label{eq:finite-time-feedback-Landauer}
  \beta Q\geq-\Delta S_{\mathrm{sys}}+I+\frac{\bigl[d_T(\rho_0,\rho_\tau)-\tau\overline b\bigr]_+^2}
  {\tau\,\overline{m}},
\end{equation}
where $[x]_+:=\max\{x,0\}$. 
% \tv{By the first law, this bound can be recast in terms of the total work done on the system, including both driving and feedback contributions.}
% By the first law, this bound can equivalently be expressed in terms of the work done on the system, including both driving and feedback work. 
Without feedback, approaching the Landauer bound generally requires the long-time limit $\tau\to\infty$~\cite{Tan20221,Liu20253}.
Remarkably, the feedback drift term in Eq.~\eqref{eq:finite-time-feedback-Landauer} reveals that when $\tau\overline b\geq d_T(\rho_0,\rho_\tau)$, the lower bound on the excess dissipation vanishes even at finite $\tau$. 
Physically, suitably chosen feedback directly drives the state transformation required for erasure, thereby reducing the component that must be driven by dissipation and suppressing entropy production.
As in experimental tests of Landauer's principle~\cite{Jun2014}, this description excludes the energy cost of the detector. The thermodynamic cost of information erasure is a physically meaningful quantity in its own right.

\begin{figure}[t]
  \centering
  \includegraphics[width=\columnwidth]{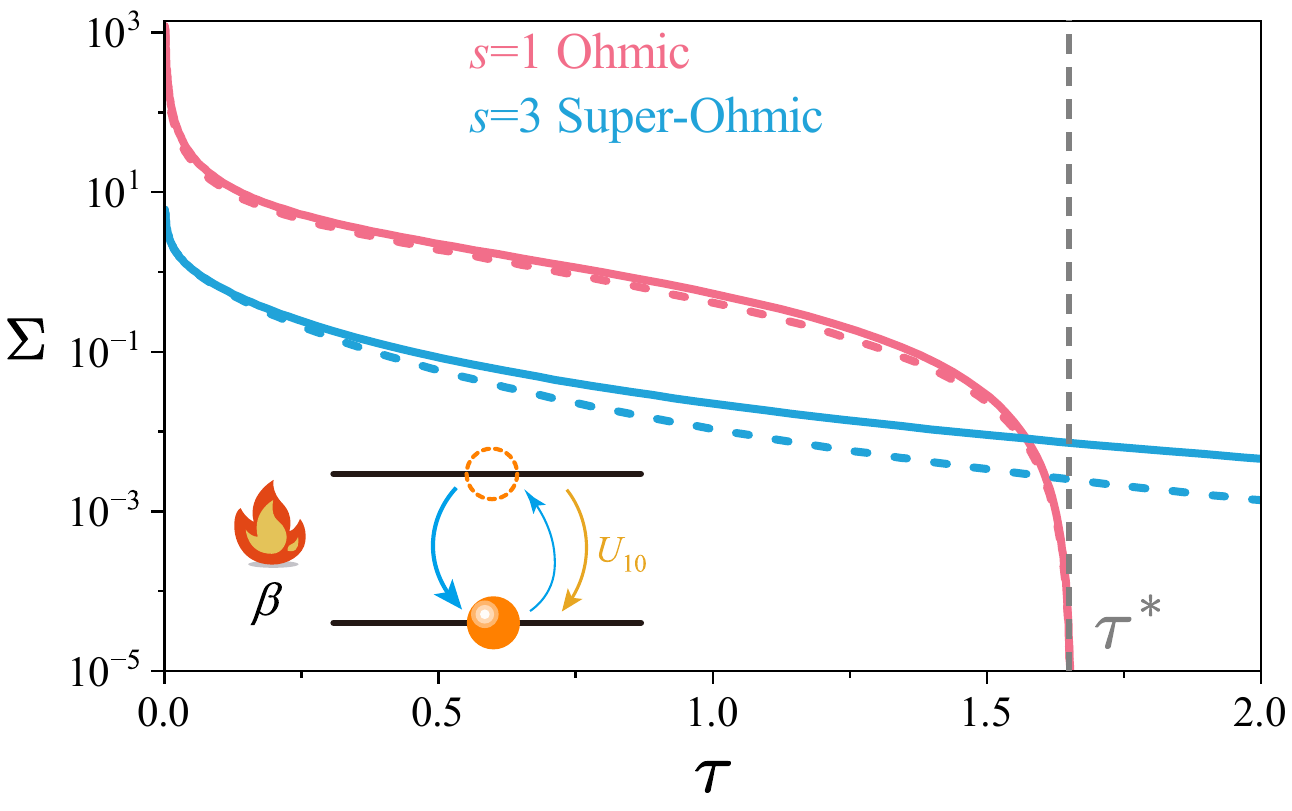}
  \caption{Finite-time information erasure with quantum feedback. Solid curves show the minimum entropy production $\Sigma_{\min}$ for $s=1$ (red) and $s=3$ (blue), with target error probability $\varepsilon=0.01$. Dashed curves show the corresponding lower bounds on entropy production, which are obtained from Eq.~\eqref{eq:finite-time-feedback-Landauer}, evaluated along the optimal protocols. The vertical gray dashed line marks the time $\tau^*$ at which $\Sigma_{\min}$ reaches zero for $s=1$. The inset shows the two-level system and the feedback $U_{10}$ applied after each jump from the ground to the excited state. Here, $\beta=1$ and $\Gamma=1$.}
  \label{fig:memory-erasure}
\end{figure}

We now illustrate the possibility of attaining the generalized Landauer bound in finite time.
We consider a two-level system (see the inset of Fig.~\ref{fig:memory-erasure}), governed by the Hamiltonian $H_t=\lambda_t\sigma_z/2$, where $\sigma_z$ is the Pauli $z$ matrix and $\lambda_t\geq0$ is the controllable energy gap, with $|0\rangle$ and $|1\rangle$ denoting the ground and excited states, respectively.
The erasure process is implemented by varying $\lambda_t$ to transform the maximally mixed state $\rho_0=\mathbb I/2$ into $\rho_\tau=(1-\varepsilon)\proj{0}+\varepsilon\proj{1}$, with target error probability $\varepsilon\ll1$. 
Previous studies on qubit reset identified coherence between energy eigenstates as a source of additional dissipation~\cite{Miller2020,Tan20221}. 
To avoid this extra dissipation, we consider a state that remains diagonal in the energy basis during the erasure process, $\rho_t=[1-p_t]\proj{0}+p_t\proj{1}$, where $p_t$ is the population of the excited state.

The jump operators $L_{0}=\sqrt{\Gamma R(\beta\lambda_t)}|0\rangle\langle1|$ and $L_{1}=\ee^{-\beta\lambda_t/2}\sqrt{\Gamma R(\beta\lambda_t)}|1\rangle\langle0|$ describe transitions from the excited to the ground state and vice versa, respectively, with $\Gamma$ denoting the coupling coefficient and $R(\beta\lambda_t)$ being the response function of the thermal reservoir. 
After each jump from the ground to the excited state, we apply a unitary feedback operation $U_{10}=|0\rangle\langle1|+|1\rangle\langle0|$ that instantaneously returns the system to the ground state (see the inset of Fig.~\ref{fig:memory-erasure}), while no feedback is applied after the reverse jump. This feedback cancels the effect of thermal excitation, yielding the population dynamics
\begin{equation}
  \dot p_t=-\Gamma p_tR(\beta\lambda_t).
\label{eq:erasure-feedback-population}
\end{equation}
To examine the thermodynamic cost of this feedback-controlled erasure for a given operation time $\tau$ and target error probability $\varepsilon$, we consider the entropy production for this process
\begin{equation}\label{eq:erasure-entropy-functional}
  \Sigma_\tau=\int_{\varepsilon}^{1/2}\dd p\,\xi(1-\ee^{-\xi}),
\end{equation}
where $\xi(p)=\beta\lambda(p)-\ln[(1-p)/p]$. 
The derivation of this expression and its minimization using the variational approach of Ref.~\cite{Liu20253} are detailed in the End Matter.

Without feedback, the population evolves increasingly slowly ($\dot p_t\to0$) in the reversible limit $\xi\to0$, requiring an infinite operation time to complete erasure. 
With feedback, however, $\dot p_t$ remains nonzero even at $\xi=0$, allowing erasure to proceed with $\Sigma_\tau=0$. 
Integrating Eq.~\eqref{eq:erasure-feedback-population} under the condition $\xi=0$, we obtain the corresponding erasure time:
\begin{equation}\label{reversible_time}
\tau^*=\frac{1}{\Gamma}\int_{\varepsilon}^{1/2}\frac{\dd p}{pR\left[\ln\frac{1-p}{p}\right]},
\end{equation}
where the reservoir response function $R$ determines whether this integral converges. 
When it converges, the generalized Landauer bound can be attained within the finite duration $\tau^*$~\cite{supp}. 
For shorter durations $\tau<\tau^*$, the minimum entropy production is obtained via the variational method~\cite{Liu20253}. 
For $\tau\geq\tau^*$, the system can remain idle for the excess duration $\tau-\tau^*$ in its initial state without increasing $\Sigma_{\tau}$ (such as decoupling from the environment), and then carry out the same erasure protocol, so that $\Sigma_{\min}(\tau)=0$.

To illustrate these findings, we consider a bosonic reservoir characterized by the response function $R(\beta\lambda_t)=\lambda_t^s/(1-\ee^{-\beta\lambda_t})$, where $s\geq0$ is the bath exponent of the spectral density. 
The integral in Eq.~\eqref{reversible_time} converges precisely for $0\leq s<2$. 
Figure~\ref{fig:memory-erasure} compares the minimum entropy production $\Sigma_{\min}$ for $s=1$ (Ohmic) and $s=3$ (super-Ohmic) with the finite-time lower bound in Eq.~\eqref{eq:finite-time-feedback-Landauer}, evaluated along the optimal protocols~\cite{supp}. 
For $s=1$, $\Sigma_{\min}$ reaches zero at the finite time $\tau^*$, demonstrating that the generalized Landauer bound is attained in finite time. 
In addition, the bound in Eq.~\eqref{eq:finite-time-feedback-Landauer} closely tracks the minimum entropy production across all operation times for both cases, including its vanishing at $\tau^*$ for $s=1$.

\textit{Power-efficiency trade-off.---}We next consider a stationary information heat engine that utilizes quantum feedback to extract work in an isothermal environment at inverse temperature $\beta$. 
This framework can be straightforwardly extended to multiple thermal reservoirs~\cite{supp}.
We choose the antisymmetric weights $c_{\nu k}^{xy}$ so that the stationary weighted current equals the output power, $J_{\rm ss}=P$.
Hereafter, the subscript ``ss'' denotes evaluation in the stationary state.
In the stationary state, the first law of thermodynamics reads $P=-\dot Q+P_F$~\cite{supp}, where $\dot Q=\beta^{-1}\sum_{\nu,k}\Delta s_{\nu k}a_{\nu k}$ is the heat current released into the environment, and $P_F=\sum_{\nu,k}a_{\nu k}\Tr\{H(\mathcal F_{\nu k}[\rho'_{\nu k}]-\rho'_{\nu k})\}$ is the net power injected into the system by the feedback controller.
The information efficiency is defined as
\begin{equation}
  \eta_I:=\frac{P}{P_F-\dot I/\beta},
\label{eq:information-engine-efficiency}
\end{equation}
where both the energy supplied to the system by the controller and the information consumed by feedback are counted as input resources~\cite{TanogamiVuSaito2023}. 
The generalized second law ensures $0\le\eta_I\leq1$ whenever $P> 0$, with $\eta_I=1$ corresponding to $\sigma_{\rm ss}=0$. 
The current--dissipation relation~\eqref{fb_cdb} then yields (see SM~\cite{supp} for details)
\begin{equation}\label{eq:information-engine-power-efficiency}
  P\leq\beta\,m_{\rm ss}\,\frac{1-\eta_I}{\eta_I},
\end{equation}
which establishes the trade-off between output power and information efficiency.  
For finite bounded $m_{\rm ss}$, the output power vanishes as $\eta_I\to1$.

%Since $\dot S_{\mathrm{sys}}=0$, Eq.~\eqref{fb_entropy_production} gives the entropy-production rate $\sigma=\beta\dot Q-\dot I=\beta P(1-\eta_I)/\eta_I\geq0$, implying $0<\eta_I\leq1$.

We make the following remarks. 
First, a trade-off of the form $(1-\eta_I)/\eta_I$ was also obtained for classical bipartite information engines~\cite{TanogamiVuSaito2023}, with the prefactor determined by long-time power fluctuations. 
Second, the weighted mobility satisfies $m_{\rm ss}\leq a_{\rm ss}/2$ (see the End Matter), where $a_{\rm ss}:=\sum_{\nu,k,x,y}(c_{\nu k}^{xy})^2 w_{\nu k}^{xy}p_y$ is the weighted dynamical activity. 
Similar weighted dynamical activity appears in power-efficiency trade-offs for classical and quantum heat engines~\cite{Shiraishi2016,Tajima2021,Funo2025}. 
Third, a sufficiently rapid divergence of $m_{\rm ss}$ leaves open the possibility of sustaining finite output power as $\eta_I\to1$. 
This requires $m_{\rm ss}$ to diverge at least as $1/(1-\eta_I)$, and hence $a_{\rm ss}$ must diverge as well.

%When the weights depend only on the monitored channel, this reduces to $a_c=\sum_{\nu,k}c_{\nu k}^2a_{\nu k}$.

%Without feedback, the population velocity contains an additional factor $1-e^{-\xi}$, which appears in the denominator of the operation-time integral~\cite{Liu20253}. The quasistatic limit $\xi\to0$, corresponding to $\Sigma\to0$, therefore requires a divergent operation time. With feedback, this factor is absent, and the reversible condition $\xi=0$ gives $\beta\lambda_{\mathrm{rev}}(p)=\ln[(1-p)/p]$ and $\Gamma\tau_{\mathrm{rev}}=\int_{\varepsilon}^{1/2} \mathrm{d}p/\{pR[\ln((1-p)/p)]\}$. Whether this integral converges is determined by $R$. Whenever it does, the generalized Landauer bound is attained in the finite time $\tau_{\mathrm{rev}}$ with $\Sigma=0$. For feasible durations $\tau<\tau_{\mathrm{rev}}$, Eqs.~\eqref{eq:erasure-optimal-protocol} and \eqref{eq:erasure-optimal-boundary} determine the optimal interior protocol and the corresponding minimum entropy production. For $\tau\geq\tau_{\mathrm{rev}}$, we allow the memory to remain decoupled from the reservoir for $\tau-\tau_{\mathrm{rev}}$ before executing the reversible protocol, giving $\Sigma_{\min}(\tau)=0$.

\begin{figure}[t]
  \centering
  \includegraphics[width=\columnwidth]{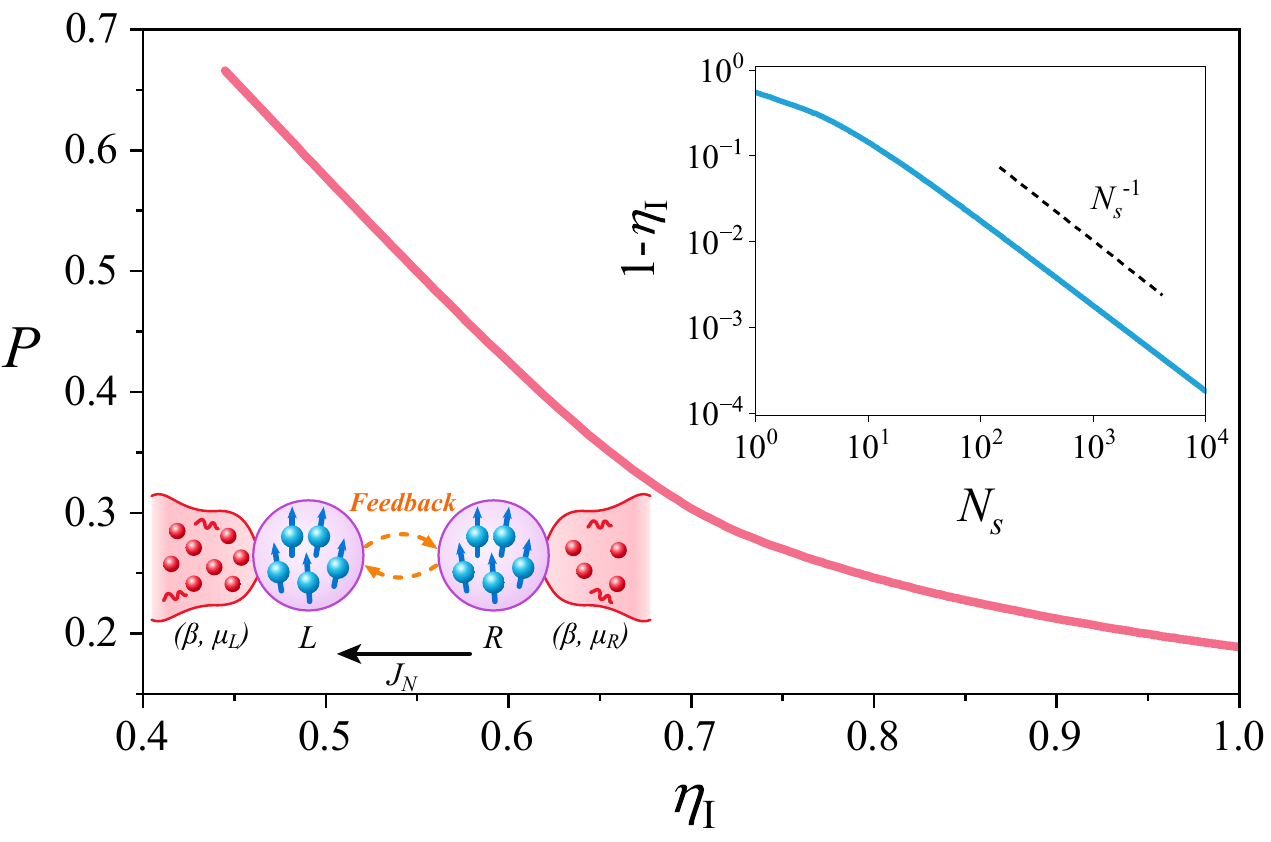}
  \caption{Power $P$ versus information efficiency $\eta_I$ for the collective information engine with increasing odd $N_s$, with $\gamma_{R,1}=1+1/N_s$. The upper-right inset shows $1-\eta_I$ versus $N_s$, with the black dashed line indicating $N_s^{-1}$ scaling. The lower-left inset illustrates the setup, where feedback drives a particle current from reservoir $R$ to reservoir $L$. Here, $\beta=\omega=\gamma_{L,1}=1$, $\mu_L=-0.5$, and $\mu_R=-1.5$.}
  \label{fig:collective-engine}
\end{figure}

\textit{Example II: Collective information heat engine.---}To demonstrate the power-efficiency trade-off, we consider an information engine that uses feedback to drive a net current against a chemical potential difference and extract chemical work.
The working medium consists of two ensembles $\alpha\in\{L,R\}$, each containing an odd number $N_s$ of identical spins collectively coupled to its own bosonic particle reservoir, as illustrated in the lower-left inset of Fig.~\ref{fig:collective-engine}.
The reservoirs share the same inverse temperature $\beta$ but different chemical potentials $\mu_L>\mu_R$. 
With $K_{\alpha,z}=\sum_{n=1}^{N_s}\sigma_{\alpha,z}^{(n)}/2$, where $\sigma_{\alpha,z}^{(n)}$ is the Pauli $z$ operator for site $n$ in ensemble $\alpha$, the Hamiltonian is
\begin{equation}\label{eq:collective-engine-Hamiltonian}
  H=\sum_{\alpha=L,R}\left(\omega K_{\alpha,z}+\frac{U}{2}K_{\alpha,z}^{2}\right),
\end{equation}
where the second term represents the interaction within each ensemble, such that the central transition has a fixed energy gap $\omega>0$, and adjacent transition energies are separated by $U$. 
This model Hamiltonian has been widely adopted to study superabsorption~\cite{Kamimura2022}.

%Collective coupling to a thermal reservoir can enhance the work output of multiqubit information engines~\cite{Jaseem2023Information}.

We consider symmetric Dicke states \(|D_n^{N_s}\rangle_\alpha\) with \(n\) excitations, which contain coherence among degenerate occupation configurations. 
By preparing each ensemble in its central Dicke doublet and coupling it to a reservoir that suppresses sideband transitions, the dynamics is effectively restricted to a two-level subspace~\cite{Kamimura2022} spanned by $|g\rangle_\alpha=|D_k^{N_s}\rangle_\alpha$ and $|e\rangle_\alpha=|D_{k+1}^{N_s}\rangle_\alpha$, with $k=(N_s-1)/2$, separated by the energy gap $\omega$. 
The two ensembles thus define four effective states: $|B\rangle=|g\rangle_L|g\rangle_R$, $|L\rangle=|e\rangle_L|g\rangle_R$, $|R\rangle=|g\rangle_L|e\rangle_R$, and $|D\rangle=|e\rangle_L|e\rangle_R$.

We normalize the collective system--reservoir coupling by $1/\sqrt{N_s}$ to ensure a well-defined large-$N_{s}$ limit, so that the central transition acquires the collective enhancement factor $\Lambda=(N_s+1)^2/(4N_s)$~\cite{supp}. 
The collective jump operators for particle injection and ejection are $L_{\alpha,1}=\sqrt{\Lambda \gamma_{\alpha,1}}\,|e\rangle_\alpha\langle g|$ and $L_{\alpha,0}=\sqrt{\Lambda \gamma_{\alpha,0}}\,|g\rangle_\alpha\langle e|$, respectively. 
Here, $\gamma_{\alpha,1}=\Gamma_\alpha n_\alpha$ and $\gamma_{\alpha,0}=\Gamma_\alpha(n_\alpha+1)$, where $n_\alpha=[\ee^{\beta(\omega-\mu_\alpha)}-1]^{-1}$ is the Bose occupation number at energy $\omega$ and $\Gamma_\alpha$ denotes the overall transition rate for ensemble $\alpha$~\cite{SantosLandi2016}. 
The rates satisfy local detailed balance, $\gamma_{\alpha,1}/\gamma_{\alpha,0}=\ee^{-\beta(\omega-\mu_\alpha)}$, with reservoir entropy changes $\Delta s_{\alpha,0}=\beta(\omega-\mu_\alpha)$ and $\Delta s_{\alpha,1}=-\beta(\omega-\mu_\alpha)$.

After detecting a particle injection from either reservoir, we apply the unitary feedback $U_{\mathrm{sw}}=\proj{B}+\proj{D} +\ketbra{L}{R}+\ketbra{R}{L}$, whereas no feedback is applied after an ejection. 
Because $|L\rangle$ and $|R\rangle$ are degenerate and contain the same number of particles, the swap conserves both energy and particle number, ensuring $P_F=0$. 
Since the jumps and feedback do not generate coherences among the four basis states, the density matrix of the steady state takes the diagonal form $\rho=p_B\proj{B}+p_L\proj{L}+p_R\proj{R}+p_D\proj{D}$.

The feedback mechanism enables particle transport between the reservoirs. 
Starting from $|B\rangle$, particle injection from reservoir $R$ prepares $|R\rangle$, which is immediately mapped to $|L\rangle$ by the feedback. 
Subsequent ejection into reservoir $L$ returns the system to $|B\rangle$, completing a cycle that transfers one particle against the chemical potential difference $\Delta\mu:=\mu_L-\mu_R>0$, thereby delivering chemical work. 
Injection from $L$ induces transport in the reverse direction. 
In the steady state, the net particle current from $R$ to $L$ reads (see the End Matter for details)
\begin{equation}
  J_N=\Lambda(\gamma_{R,1}-\gamma_{L,1})p_B,
\label{eq:collective-engine-current}
\end{equation}
which suggests that the information heat engine requires $\gamma_{R,1}>\gamma_{L,1}$. 
Without feedback, each ensemble equilibrates with its own reservoir and the particle current vanishes.

The information term associated with the feedback reads $\dot{I}= J_N\ln(p_R/p_L)<0$, which indicates information consumption (see the End Matter). 
Combining with $P_F=0$, the corresponding information efficiency and output power are written as
\begin{equation}\label{eq:collective-engine-efficiency-power}
  \eta_I=\frac{\Delta\mu}{\ln(p_L/p_R)/\beta},\qquad P=\Delta\mu\,J_N.
\end{equation}
With the weighted mobility $m_{\rm ss}$ given in the End Matter, the power-efficiency trade-off in Eq.~\eqref{eq:information-engine-power-efficiency} is exactly saturated.

Although Eq.~\eqref{eq:information-engine-power-efficiency} dictates that output power generally vanishes as $\eta_I\to1$ for finite $m_{\rm ss}$, a diverging mobility can circumvent this trade-off. 
To realize this possibility, we keep the bath parameters and $\gamma_{L,1}$ fixed and choose $\gamma_{R,1}=\gamma_{L,1}+1/N_s$. 
Consequently, the collective enhancement yields a linearly diverging mobility $m_{\rm ss}\propto N_s$. 
In the limit $N_s\to\infty$, we obtain~\cite{supp}
\begin{equation}\label{eq:collective-engine-unit-efficiency-limit}
  1-\eta_I=O(N_s^{-1}),\qquad P\longrightarrow\frac{\Delta\mu}{4}p_B^{\mathrm{rev}}>0,
\end{equation}
where $p_B^{\mathrm{rev}}=\prod_{\alpha=L,R}[1+\ee^{-\beta(\omega-\mu_\alpha)}]^{-1}$ is the limiting population at $\gamma_{R,1}=\gamma_{L,1}$. 
As indicated by Eq.~\eqref{eq:collective-engine-current}, the collective enhancement ($\Lambda\propto N_s$) compensates for the vanishing difference between the injection rates ($\gamma_{R,1}-\gamma_{L,1}\propto N_s^{-1}$), sustaining a finite output power while $\sigma_{\rm ss}\to0$. 
Figure~\ref{fig:collective-engine} shows that $P$ approaches a nonzero value as $\eta_I\to1$, while the inset confirms the $N_s^{-1}$ scaling of $1-\eta_I$.

\textit{Summary.---}To clarify how quantum feedback reshapes finite-time thermodynamic bounds, we established a fundamental finite-time current--dissipation relation for open quantum dynamics with feedback, from which we derived the quantum speed limit for state transformations and the power-efficiency trade-off for stationary information engines.
These results reveal two complementary roles of feedback: information modifies the entropy balance, while feedback operations directly reshape the system dynamics.
In finite-time information erasure, direct feedback driving enables the generalized Landauer bound to be saturated within \emph{finite} time, while in a collective information engine, many-body cooperative effects allow unit information efficiency to be approached at finite output power. 
%Our findings provide a framework for optimizing speed and dissipation in feedback-controlled quantum technologies.

\begin{acknowledgements}
\textit{Acknowledgments.}---YL gratefully acknowledges the Yukawa Research Fellow, co-sponsored by the YITP and the Yukawa Memorial Foundation.
TVV was supported by JSPS KAKENHI Grant Nos.~JP23K13032, JP26K00019, JP26K00022, and JP26H02015.
RB was supported by JSPS KAKENHI Grant No. JP25KJ0766.
\end{acknowledgements}

\textit{Data availability.}---The data are not publicly available upon publication. The data are available from the authors upon reasonable request.

\bibliography{References}

\section*{End Matter}
\appendix
\renewcommand{\theHequation}{\thesection.\arabic{equation}}

\refstepcounter{section}
\textit{Appendix \thesection: Weighted dynamical state mobility.---}In the instantaneous eigenbasis of $\rho_t$, the transition rate from $|y_t\rangle$ to $|x_t\rangle$ through the channel $(\nu,k)$ is $w_{\nu k}^{xy}(t):=|\langle x_t|L_{\nu k}|y_t\rangle|^2$. 
The probability current and thermodynamic force for each pair of forward and reverse transitions are
\begin{equation}\label{eq:end-matter-current-force}
\begin{aligned}
  j_{\nu k}^{xy}&:=w_{\nu k}^{xy}p_y-w_{\nu k^*}^{yx}p_x,\\
  f_{\nu k}^{xy}&:=\ln\frac{w_{\nu k}^{xy}p_y}{w_{\nu k^*}^{yx}p_x}.
\end{aligned}
\end{equation}
Local detailed balance gives $f_{\nu k}^{xy}=\ln(p_y/p_x)+\Delta s_{\nu k}$. 
The logarithmic mean of the two probability fluxes is $\ell_{\nu k}^{xy}:=j_{\nu k}^{xy}/f_{\nu k}^{xy}\geq0$, defined by continuity as $\ell_{\nu k}^{xy}=w_{\nu k}^{xy}p_y$ when the fluxes coincide. 
The weighted mobility is then
\begin{equation}
  m_t:=\frac12\sum_{\nu,k,x,y}
  [c_{\nu k}^{xy}(t)]^2\ell_{\nu k}^{xy}(t).
\label{eq:end-matter-general-coefficient}
\end{equation}
Since both $c$ and $j$ change sign under reversal, the weighted current can equivalently be written as 
\begin{equation}
  J_t=\frac12\sum_{\nu,k,x,y}c_{\nu k}^{xy}j_{\nu k}^{xy}=\sum_{\nu,k,x,y}c_{\nu k}^{xy}w_{\nu k}^{xy}p_y,
\end{equation}
which is consistent with Eq.~\eqref{eq:weighted-current-coefficient}.

By using $(x-y)\ln(x/y)\geq 2(x-y)^{2}/(x+y)$, Eq.~\eqref{eq:end-matter-general-coefficient} therefore gives
\begin{equation}\label{eq:end-matter-activity-bound}
\begin{aligned}
  m_t&\leq\frac14\sum_{\nu,k,x,y}(c_{\nu k}^{xy})^2\bigl(w_{\nu k}^{xy}p_y+w_{\nu k^*}^{yx}p_x\bigr)\\
  &=\frac12 \sum_{\nu,k,x,y}(c_{\nu k}^{xy})^2w_{\nu k}^{xy}p_y.
\end{aligned}
\end{equation}
The last equality follows by relabeling the reverse transitions and using $(c_{\nu k^*}^{yx})^2=(c_{\nu k}^{xy})^2$.

For the quantum speed limit in Eq.~\eqref{fb_qsl}, the weights $c_{\nu k}^{xy}=z_x-z_y$, with $z_x=\operatorname{sgn}[p_x(\tau)-p_x(0)]$, give
\begin{equation}\label{eq:end-matter-qsl-coefficient}
  m_t=\frac12\sum_{\nu,k,x,y}(z_x-z_y)^2\ell_{\nu k}^{xy}(t).
\end{equation}
In this case, Eq.~\eqref{eq:end-matter-activity-bound} and $|z_x-z_y|\leq2$ imply $m_t\leq 2a_t$, where $a_t=\sum_{\nu,k,x,y}w_{\nu k}^{xy}p_y$ is the dynamical activity.

For the power-efficiency trade-off in Eq.~\eqref{eq:information-engine-power-efficiency}, the weights are chosen so that $J_t=P$ in the stationary state. 
For example, in the collective information engine, we may assign $c_{R,1}^{xy}=\Delta\mu$ to particle injection from reservoir $R$, $c_{R,0}^{xy}=-\Delta\mu$ to ejection into that reservoir, and $c_{L,k}^{xy}=0$ to the transitions induced by reservoir $L$. 
These weights give $J_{\rm ss}=\Delta\mu\,J_N=P$ in the steady state.
An optimized choice of weights and the corresponding weighted mobility are given in Eq.~\eqref{eq:end-matter-engine-coefficient}.\\

\refstepcounter{section}
\textit{Appendix \thesection: Entropy production for the information erasure process.---}We first derive the entropy production in Eq.~\eqref{eq:erasure-entropy-functional}. 
For the erasure model, the feedback master equation~\eqref{FB_Lindblad} takes the form
\begin{equation}\label{eq:erasure-feedback-master}
  \dot\rho_t=-i[H_t,\rho_t]+\Diss{L_0}[\rho_t]+\Diss{U_{10}L_1}[\rho_t],
\end{equation}
where $\Diss{X}[\rho]:=X\rho X^{\dagger}-\{X^{\dagger}X,\rho\}/2$. 
The last term follows from $U_{10}\ketbra{1}{0}=\proj{0}$.

For the diagonal state $\rho_t=(1-p_t)\proj{0}+p_t\proj{1}$, the bare jump rates are $a_0=\Gamma R(\beta\lambda_t)p_t$ and $a_1=\Gamma R(\beta\lambda_t)\ee^{-\beta\lambda_t}(1-p_t)$. 
Both the Hamiltonian term and the last dissipator in Eq.~\eqref{eq:erasure-feedback-master} vanish for this state, giving $\dot p_t=-a_0$ as in Eq.~\eqref{eq:erasure-feedback-population}.

The rate of change of the system entropy reads $\dot S_{\mathrm{sys}}=\dot p_t\ln[(1-p_t)/p_t]=-a_0\ln[(1-p_t)/p_t]$, and the environment entropy change is given by $\dot S_{\mathrm{env}}=\beta\lambda_t(a_0-a_1)$.
Only the feedback following an upward jump contributes to the information term in Eq.~\eqref{information}. 
It maps the conditional state $\proj{1}$ to $\proj{0}$. 
Using $D(\proj{0}\Vert\rho_t)=-\ln(1-p_t)$ and $D(\proj{1}\Vert\rho_t)=-\ln p_t$, we obtain
\begin{equation}\label{eq:erasure-information-rate}
\begin{aligned}
  \dot I&=a_1\bigl[D(\proj{0}\Vert\rho_t)-D(\proj{1}\Vert\rho_t)\bigr]\\
  &=a_1\ln\frac{p_t}{1-p_t}.
\end{aligned}
\end{equation}
The rate of entropy production therefore reads
\begin{equation}
\begin{aligned}
  \sigma_t
  &=\dot S_{\mathrm{sys}}+\dot S_{\mathrm{env}}-\dot I\\
  &=(a_0-a_1)\left[\beta\lambda_t-\ln\frac{1-p_t}{p_t}\right]\\
  &=a_0\xi(p_t)(1-\ee^{-\xi(p_t)}).
\end{aligned}
\label{eq:erasure-entropy-rate}
\end{equation}
Integrating this rate with $\mathrm{d}t=-\mathrm{d}p_t/a_0$ from $p_0=1/2$ to $p_\tau=\varepsilon$ yields Eq.~\eqref{eq:erasure-entropy-functional}.

We next minimize this functional for the given operation time $\tau$ and the set error probability $\varepsilon$. 
Following the variational approach in Ref.~\cite{Liu20253}, the optimal protocol $\lambda(p)$ satisfies~\cite{supp}
\begin{equation}\label{eq:erasure-optimal-protocol}
  CpR(\beta\lambda)\bigl(1-\ee^{-\xi}+\xi \ee^{-\xi}\bigr)-\frac{R'(\beta\lambda)}{R(\beta\lambda)}=0,
\end{equation}
where the constant $C$ is determined by the boundary conditions $p_0=1/2$ and $p_\tau=\varepsilon$. 
Integrating Eq.~\eqref{eq:erasure-feedback-population} gives
\begin{equation}\label{eq:erasure-optimal-boundary}
  \Gamma\tau=\int_{\varepsilon}^{1/2}\frac{\dd p}{pR[\beta\lambda(p)]}.
\end{equation}
Substituting the optimal solution $\lambda(p)$ obtained from Eqs.~\eqref{eq:erasure-optimal-protocol} and \eqref{eq:erasure-optimal-boundary} into Eq.~\eqref{eq:erasure-entropy-functional} yields the minimum entropy production $\Sigma_{\min}(\tau)$.\\

\refstepcounter{section}
\textit{Appendix \thesection: Steady-state properties of the information engine.---}For the collective information engine, the feedback master equation~\eqref{FB_Lindblad} becomes
\begin{equation}\label{eq:engine-feedback-master}
\begin{aligned}
  \dot\rho_t={}&-i[H,\rho_t]\\
  &+\sum_{\alpha=L,R}\Bigl(
  \Diss{L_{\alpha,0}}[\rho_t]+\Diss{U_{\mathrm{sw}}L_{\alpha,1}}[\rho_t]\Bigr).
\end{aligned}
\end{equation}
The swap $U_{\mathrm{sw}}$ is applied after particle injection from either reservoir.
For a state diagonal in the basis $\{|B\rangle,|L\rangle,|R\rangle,|D\rangle\}$, Eq.~\eqref{eq:engine-feedback-master} gives
\begin{equation}\label{eq:engine-population-dynamics}
\begin{aligned}
\dot p_B
&=\Lambda\bigl[-(\gamma_{L,1}+\gamma_{R,1})p_B
+\gamma_{L,0}p_L+\gamma_{R,0}p_R\bigr],\\
\dot p_L
&=\Lambda\bigl[\gamma_{R,1}p_B-(\gamma_{L,0}+\gamma_{R,1})p_L
+\gamma_{R,0}p_D\bigr],\\
\dot p_R
&=\Lambda\bigl[\gamma_{L,1}p_B-(\gamma_{R,0}+\gamma_{L,1})p_R
+\gamma_{L,0}p_D\bigr],\\
\dot p_D
&=\Lambda\bigl[\gamma_{R,1}p_L+\gamma_{L,1}p_R
-(\gamma_{L,0}+\gamma_{R,0})p_D\bigr].
\end{aligned}
\end{equation}
Defining $\gamma_k=\gamma_{L,k}+\gamma_{R,k}$, the stationary solution can be written as
\begin{equation}\label{eq:engine-stationary-populations}
\begin{aligned}
p_L&=\frac{\gamma_{R,1}\gamma_0+\gamma_{L,1}\gamma_1}
{\gamma_{L,0}(\gamma_0+\gamma_1)}p_B,\\
p_R&=\frac{\gamma_{L,1}\gamma_0+\gamma_{R,1}\gamma_1}
{\gamma_{R,0}(\gamma_0+\gamma_1)}p_B,\\
p_D&=\frac{\gamma_{R,1}p_L+\gamma_{L,1}p_R}{\gamma_0},
\end{aligned}
\end{equation}
where $p_B$ is fixed by normalization. 
The net particle current from reservoir $R$ can be expressed as
\begin{equation}
\begin{aligned}
J_N&=\Lambda\bigl[\gamma_{R,1}(p_B+p_L)
-\gamma_{R,0}(p_R+p_D)\bigr]\\
&=\Lambda(\gamma_{R,1}-\gamma_{L,1})p_B,
\end{aligned}
\label{eq:engine-current-derivation}
\end{equation}
where the second equality follows from the stationary condition $\dot p_R=0$ and $\dot p_D=0$. 
This gives Eq.~\eqref{eq:collective-engine-current}.

We next evaluate the information term in Eq.~\eqref{information}. 
For particle injection from reservoir $R$, the jump rate is $a_{R,1}=\Lambda\gamma_{R,1}(p_B+p_L)$, and the conditional state before feedback is $\rho'_{R,1}=(p_B\proj{R}+p_L\proj{D})/(p_B+p_L)$. 
The feedback maps $|R\rangle$ to $|L\rangle$ and leaves $|D\rangle$ unchanged. 
Since the unitary operation preserves the entropy of the conditional state, Eq.~\eqref{information} gives the contribution $\Lambda\gamma_{R,1}p_B\ln(p_R/p_L)$ per unit time. 
The contribution from particle injection at reservoir $L$ follows by exchanging $L$ and $R$. 
Ejections do not contribute because no feedback is applied. 
The total information rate therefore reads
\begin{equation}\label{eq:engine-information-derivation}
\begin{aligned}
\dot I&=\Lambda(\gamma_{R,1}-\gamma_{L,1})p_B\ln\frac{p_R}{p_L}\\
&=J_N\ln\frac{p_R}{p_L}.
\end{aligned}
\end{equation}

To evaluate the weighted mobility, we choose the weights $c_{\alpha k}^{xy}=\Delta\mu f_{\alpha k}^{xy}/[\ln(p_L/p_R)-\beta\Delta\mu]$~\cite{supp}. 
The stationary solution gives $\frac12\sum_{\alpha,k,x,y}j_{\alpha k}^{xy}f_{\alpha k}^{xy}=J_N[\ln(p_L/p_R)-\beta\Delta\mu]$, so that the weighted current equals the output power, $J_{\rm ss}=P$. 
Substituting them into Eq.~\eqref{eq:end-matter-general-coefficient} yields
\begin{equation}\label{eq:end-matter-engine-coefficient}
\begin{aligned}
  m_{\rm ss}&=(\Delta\mu)^2\left(\frac{\ell_{L,B}\ell_{R,B}}{\ell_{L,B}+\ell_{R,B}}+\frac{\ell_{L,D}\ell_{R,D}}{\ell_{L,D}+\ell_{R,D}}\right)\\
  &=\frac{(\Delta\mu)^2J_N}{\ln(p_L/p_R)-\beta\Delta\mu}.
\end{aligned}
\end{equation}
Here, $\ell_{\alpha,B}$ and $\ell_{\alpha,D}$ denote the logarithmic means of the forward and reverse probability fluxes for the transitions induced by reservoir $\alpha$ involving $|B\rangle$ and $|D\rangle$, respectively. 
These fluxes include the collective factor $\Lambda$. 
Together with Eq.~\eqref{eq:collective-engine-efficiency-power}, this gives $P=\beta m_{\rm ss}(1-\eta_I)/\eta_I$, showing that Eq.~\eqref{eq:information-engine-power-efficiency} is saturated.

\end{document}

\title{Supplemental Material for\\
``Thermodynamics of Speed and Efficiency with Quantum Feedback''}

\author{Yue Liu}
\email{yue.liu@yukawa.kyoto-u.ac.jp}
\affiliation{Center for Gravitational Physics and Quantum Information, Yukawa Institute for Theoretical Physics, Kyoto University, Kitashirakawa Oiwakecho, Sakyo-ku, Kyoto 606-8502, Japan}

\author{Ryotaro Honma}
\affiliation{Center for Gravitational Physics and Quantum Information, Yukawa Institute for Theoretical Physics, Kyoto University, Kitashirakawa Oiwakecho, Sakyo-ku, Kyoto 606-8502, Japan}

\author{Ruicheng Bao}
\affiliation{Department of Physics, Graduate School of Science, The University of Tokyo, Hongo, Bunkyo-ku, Tokyo 113-0033, Japan}

\author{Tan Van Vu}
\email{tan.vu@yukawa.kyoto-u.ac.jp}
\affiliation{Center for Gravitational Physics and Quantum Information, Yukawa Institute for Theoretical Physics, Kyoto University, Kitashirakawa Oiwakecho, Sakyo-ku, Kyoto 606-8502, Japan}

\maketitle

\tableofcontents

\section{Wiseman master equation}
\label{sec:feedback-dynamics}

We consider a quantum system weakly coupled to thermal reservoirs $\nu$ at inverse temperatures $\beta_\nu$. 
Under continuous monitoring, a detected jump $(\nu,k)$ is followed immediately by a feedback operation $\cF_{\nu k}$. 
For ideal detection and negligible feedback delay, the density matrix obeys the Wiseman master equation,
\begin{equation}\label{eq:wiseman}
  \dot\rho_t=-i[H_t,\rho_t]+\sum_{\nu,k}\left(\cF_{\nu k}[L_{\nu k}\rho_tL_{\nu k}^{\dagger}]-\frac12\{L_{\nu k}^{\dagger}L_{\nu k},\rho_t\}\right)\equiv\cL_t^{(\mathrm{fb})}[\rho_t].
\end{equation}
Here, $H_t$ is the system Hamiltonian and $L_{\nu k}$ describes a jump in channel $k$ induced by reservoir $\nu$. 
We set $\hbar=k_{\mathrm B}=1$ for simplicity. 
The jump operators and feedback operations may depend on time, although their explicit time labels are suppressed. 
As in the main text, each $\cF_{\nu k}$ is a completely positive, trace-preserving, and unital map,
\begin{equation}\label{eq:feedback-map}
  \cF_{\nu k}[X]=\sum_\alpha F_{\nu k\alpha}X F_{\nu k\alpha}^{\dagger},\qquad \sum_\alpha F_{\nu k\alpha}^{\dagger}F_{\nu k\alpha}=\sum_\alpha F_{\nu k\alpha}F_{\nu k\alpha}^{\dagger}=\mathbb I.
\end{equation}
This class includes unitary transformations, nonselective projective measurements, and the identity map.

The bare jump operators satisfy local detailed balance,
\begin{equation}\label{eq:local-detailed-balance}
  L_{\nu k}=\ee^{\Delta s_{\nu k}/2}L_{\nu k^*}^{\dagger},\qquad\Delta s_{\nu k^*}=-\Delta s_{\nu k},
\end{equation}
where $k^*$ labels the reverse channel and $\Delta s_{\nu k}$ is the reservoir entropy change associated with the jump $L_{\nu k}$. 
The reservoir entropy change is assigned to the bare jump, before the feedback operation is applied.

The jump rate and the normalized conditional state before feedback are
\begin{equation}\label{eq:conditional-state}
  a_{\nu k}(t):=\Tr(L_{\nu k}\rho_tL_{\nu k}^{\dagger}),\qquad\rho'_{\nu k,t}:=\frac{L_{\nu k}\rho_tL_{\nu k}^{\dagger}}{a_{\nu k}(t)}.
\end{equation}
The state after feedback is $\cF_{\nu k}[\rho'_{\nu k,t}]$. 
Trace preservation implies that the feedback operation does not change the probability assigned to the detected jump. 
Channels with $a_{\nu k}(t)=0$ give no contribution and are omitted when conditional states are used.

To separate the direct contribution of feedback to the dynamics, we write
\begin{equation}\label{eq:split-dynamics}
  \dot\rho_t=\cL_t^0[\rho_t]+\cB_t[\rho_t],
\end{equation}
where
\begin{align}
  \cL_t^0[\rho]&:=-i[H_t,\rho]+\sum_{\nu,k}\Diss{L_{\nu k}}[\rho],\qquad\Diss{L}[\rho]:=L\rho L^{\dagger}-\frac12\{L^{\dagger}L,\rho\},
  \label{eq:bare-generator}\\
  \cB_t[\rho]&:=\sum_{\nu,k}\left(\cF_{\nu k}[L_{\nu k}\rho L_{\nu k}^{\dagger}]-L_{\nu k}\rho L_{\nu k}^{\dagger}\right).
  \label{eq:feedback-drift}
\end{align}
In terms of the conditional states, the feedback contribution becomes
\begin{equation}\label{eq:conditional-drift}
  \cB_t[\rho_t]=\sum_{\nu,k}a_{\nu k}(t)\left(\cF_{\nu k}[\rho'_{\nu k,t}]-\rho'_{\nu k,t}\right).
\end{equation}
Both terms in Eq.~\eqref{eq:split-dynamics} are evaluated on the actual
state $\rho_t$ evolving under feedback.

\section{Current--dissipation relation}
\label{sec:current-dissipation}

We first derive the entropy production before and after feedback, then use their comparison to establish the current--dissipation bound.
Here, before and after refer to the conditional feedback operation within each infinitesimal monitoring step. 
All quantities are evaluated along the actual feedback trajectory $\rho_t$.

\subsection{Entropy production before feedback}
\label{sec:bare-entropy-production}

The entropy production before feedback is associated with the bare generator $\cL_t^0$. 
To express it in terms of transition currents, we consider the instantaneous eigenbasis of $\rho_t$, $\rho_t=\sum_y p_y(t)\proj{y_t}$, with $p_y(t)>0$. 
In this basis,
\begin{equation}\label{eq:transition-rate}
  w_{\nu k}^{xy}(t):=|\langle x_t|L_{\nu k}|y_t\rangle|^2
\end{equation}
is the transition rate from $|y_t\rangle$ to $|x_t\rangle$ through channel $(\nu,k)$, and $w_{\nu k}^{xy}p_y$ is the corresponding probability flux. 
The reverse transition has flux $w_{\nu k^*}^{yx}p_x$. The probability current and thermodynamic force are
\begin{equation}\label{eq:current-force}
  j_{\nu k}^{xy}:=w_{\nu k}^{xy}p_y-w_{\nu k^*}^{yx}p_x,\qquad f_{\nu k}^{xy}:=\ln\frac{w_{\nu k}^{xy}p_y}{w_{\nu k^*}^{yx}p_x}=\ln\frac{p_y}{p_x}+\Delta s_{\nu k},
\end{equation}
where the last equality follows from local detailed balance, $w_{\nu k}^{xy}=\ee^{\Delta s_{\nu k}}w_{\nu k^*}^{yx}$. 
The logarithmic mean of the forward and reverse fluxes reads
\begin{equation}\label{eq:logarithmic-mean}
  \ell_{\nu k}^{xy}:=\frac{w_{\nu k}^{xy}p_y-w_{\nu k^*}^{yx}p_x}{\ln(w_{\nu k}^{xy}p_y)-\ln(w_{\nu k^*}^{yx}p_x)}.
\end{equation}
When the forward and reverse probability fluxes are equal, this expression is defined by continuity as $\ell_{\nu k}^{xy}=w_{\nu k}^{xy}p_y$. 
An inactive pair has $\ell_{\nu k}^{xy}=0$ and is omitted from sums involving the force or a quotient by $\ell_{\nu k}^{xy}$. 
On every active pair, the two fluxes are positive and $\ell_{\nu k}^{xy}>0$. These quantities satisfy
\begin{equation}\label{eq:reversal-and-force-current}
  j_{\nu k^*}^{yx}=-j_{\nu k}^{xy},
  \qquad f_{\nu k^*}^{yx}=-f_{\nu k}^{xy},
  \qquad \ell_{\nu k^*}^{yx}=\ell_{\nu k}^{xy},
  \qquad j_{\nu k}^{xy}=\ell_{\nu k}^{xy}f_{\nu k}^{xy}.
\end{equation}

With the von Neumann entropy $S(\rho):=-\Tr(\rho\ln\rho)$, the bare entropy production rate is defined as
\begin{equation}
  \sigma_t^{(0)}:=-\Tr\!\left(\cL_t^0[\rho_t]\ln\rho_t\right)
  +\sum_{\nu,k}\Delta s_{\nu k}(t)a_{\nu k}(t).
  \label{eq:bare-entropy-production}
\end{equation}
The superscript $(0)$ refers to the bare generator evaluated on $\rho_t$. 
The Hamiltonian term does not contribute to Eq.~\eqref{eq:bare-entropy-production}, since $\Tr([H_t,\rho_t]\ln\rho_t)=\Tr(H_t[\rho_t,\ln\rho_t])=0$. 
For each dissipative term, expansion in the instantaneous eigenbasis yields
\begin{equation}\label{eq:dissipative-entropy}
  \begin{split}
    -\Tr\!\left(\Diss{L_{\nu k}}[\rho_t]\ln\rho_t\right)&=-\sum_{x,y}w_{\nu k}^{xy}p_y\ln p_x+\sum_y p_y\ln p_y\langle y_t|L_{\nu k}^{\dagger}L_{\nu k}|y_t\rangle\\
    &=\sum_{x,y}w_{\nu k}^{xy}p_y\ln\frac{p_y}{p_x}.
  \end{split}
\end{equation}
Using $a_{\nu k}=\sum_{x,y}w_{\nu k}^{xy}p_y$ and Eq.~\eqref{eq:current-force}, we obtain
\begin{equation}\label{eq:bare-directed-sum}
  \sigma_t^{(0)}=\sum_{\nu,k,x,y}w_{\nu k}^{xy}p_y\left(\ln\frac{p_y}{p_x}+\Delta s_{\nu k}\right)=\sum_{\nu,k,x,y}w_{\nu k}^{xy}p_y f_{\nu k}^{xy}.
\end{equation}
Pairing each transition with its reverse gives
\begin{equation}\label{eq:bare-nonnegative}
  \sigma_t^{(0)}=\frac12\sum_{\nu,k,x,y}j_{\nu k}^{xy}f_{\nu k}^{xy}=\frac12\sum_{\nu,k,x,y}\frac{(j_{\nu k}^{xy})^2}{\ell_{\nu k}^{xy}}=\frac12\sum_{\nu,k,x,y}\ell_{\nu k}^{xy}(f_{\nu k}^{xy})^2\geq0.
\end{equation}
The factor $1/2$ accounts for summing over both members of each reverse pair.
Nonnegativity follows from $\ell_{\nu k}^{xy}\geq0$ on each pair. 
Equality holds precisely when the forward and reverse probability fluxes coincide on every pair. 
The integrated entropy production is therefore nonnegative,
\begin{equation}\label{eq:integrated-bare-entropy-production}
  \Sigma_\tau^{(0)}:=\int_0^\tau\sigma_t^{(0)}\,\dd t\geq0.
\end{equation}

\subsection{Entropy production after feedback}
\label{sec:entropy-production}

The entropy production after feedback includes the information contribution of the conditional operation. 
We derive this contribution from the change in mutual information between the system and the measurement record. 

Consider an infinitesimal interval $\mathrm{d}t$. 
A jump $(\nu,k)$ occurs with probability $q_{\nu k}=a_{\nu k}(t)\,\mathrm{d}t+O(\mathrm{d}t^2)$, and the probability of no jump is $q_0=1-a_t\,\mathrm{d}t+O(\mathrm{d}t^2)$, where $a_t:=\sum_{\nu,k}a_{\nu k}(t)$. 
The no-jump state is
\begin{equation}\label{eq:no-jump-state}
  \rho_{\mathrm{nj}}=\frac{M_0\rho_tM_0^{\dagger}}{\Tr(M_0\rho_tM_0^{\dagger})},\qquad M_0=\mathbb I-\left(iH_t+\frac12\sum_{\nu,k}L_{\nu k}^{\dagger}L_{\nu k}\right)\,\dd t+O(\dd t^2).
\end{equation}
No feedback is applied to this outcome. 
Let a classical register $M$ store the detected outcome in orthogonal states $|0\rangle_M$ and $|\nu,k\rangle_M$. 
To first order in $\mathrm{d}t$, the joint states before and after feedback are
\begin{align}
  \varrho_{SM}^{-}&=q_0\rho_{\mathrm{nj}}\otimes\proj{0}_M+\sum_{\nu,k}q_{\nu k}\rho'_{\nu k,t}\otimes\proj{\nu,k}_M,
  \label{eq:record-before}\\
  \varrho_{SM}^{+}&=q_0\rho_{\mathrm{nj}}\otimes\proj{0}_M+\sum_{\nu,k}q_{\nu k}\cF_{\nu k}[\rho'_{\nu k,t}]\otimes\proj{\nu,k}_M.
  \label{eq:record-after}
\end{align}
Tracing out the record gives $\bar\rho^\pm:=\Tr_M\varrho_{SM}^{\pm}$, the system states averaged over all detection outcomes before ($-$) and after ($+$) feedback. 
They satisfy
\begin{equation}\label{eq:record-marginals}
  \bar\rho^{-}=\rho_t+\cL_t^0[\rho_t]\dd t+O(\dd t^2),\qquad
  \bar\rho^{+}=\rho_t+\bigl(\cL_t^0[\rho_t]+\cB_t[\rho_t]\bigr)\dd t+O(\dd t^2).
\end{equation}

The quantum mutual information is defined as $\mathcal I(S{:}M):=S(\Tr_M\varrho_{SM})+S(\Tr_S\varrho_{SM})-S(\varrho_{SM})$.
For a classical record with probabilities $q_z$ and conditional system states $\rho_z$, it becomes
\begin{equation}
  \mathcal I(S{:}M)
  =S(\bar\rho)-\sum_zq_zS(\rho_z)
  =\sum_zq_zD(\rho_z\Vert\bar\rho)\geq0,
  \label{eq:mutual-information}
\end{equation}
where $D(\rho\Vert\chi):=\Tr[\rho(\ln\rho-\ln\chi)]$ is the quantum relative entropy.
We denote the mutual information of $\varrho_{SM}^\pm$ by $\mathcal I^\pm$.
Feedback preserves the record probabilities and leaves the no-jump conditional state unchanged. 
Its contribution to the mutual-information change is therefore
\begin{equation}\label{eq:mutual-information-change}
  \mathcal I^{+}-\mathcal I^{-}=S(\bar\rho^{+})-S(\bar\rho^{-})-\sum_{\nu,k}q_{\nu k}\left[S(\cF_{\nu k}[\rho'_{\nu k,t}])-S(\rho'_{\nu k,t})\right]+o(\dd t).
\end{equation}
Using Eq.~\eqref{eq:record-marginals} and $\Tr\cB_t[\rho_t]=0$, the change in marginal entropy reads
\begin{equation}\label{eq:marginal-entropy-change}
  S(\bar\rho^{+})-S(\bar\rho^{-})=-\Tr\!\left(\cB_t[\rho_t]\ln\rho_t\right)\dd t+O(\dd t^2).
\end{equation}
Consequently, the information rate associated with feedback is
\begin{equation}\label{eq:information-entropy-form}
  \dot I(t):=\lim_{\dd t\to0}\frac{\mathcal I^{+}-\mathcal I^{-}}{\dd t}=-\Tr\!\left(\cB_t[\rho_t]\ln\rho_t\right)-\sum_{\nu,k}a_{\nu k}(t)\left[S(\cF_{\nu k}[\rho'_{\nu k,t}])-S(\rho'_{\nu k,t})\right].
\end{equation}
For each conditional state, the relative entropy difference satisfies
\begin{equation}\label{eq:relative-entropy-difference}
  D(\cF_{\nu k}[\rho'_{\nu k,t}]\Vert\rho_t)-D(\rho'_{\nu k,t}\Vert\rho_t)=-S(\cF_{\nu k}[\rho'_{\nu k,t}])+S(\rho'_{\nu k,t})-\Tr\!\left[(\cF_{\nu k}[\rho'_{\nu k,t}]-\rho'_{\nu k,t})\ln\rho_t\right].
\end{equation}
Substituting Eq.~\eqref{eq:conditional-drift} into Eq.~\eqref{eq:information-entropy-form} thus gives
\begin{equation}\label{eq:information-rate}
  \dot I(t)=\sum_{\nu,k}a_{\nu k}(t)\left[D(\cF_{\nu k}[\rho'_{\nu k,t}]\Vert\rho_t)-D(\rho'_{\nu k,t}\Vert\rho_t)\right],
  \qquad
  I=\int_0^\tau\dot I(t)\,\dd t,
\end{equation}
which is the expression used in the main text. 
The information term is defined as the change after feedback minus before feedback. 
It can have either sign, with $I<0$ corresponding to information consumption.

The rates of change in system and reservoir entropy are
\begin{equation}\label{eq:entropy-rates}
  \dot S_{\mathrm{sys}}(t)=-\Tr(\dot\rho_t\ln\rho_t),
  \qquad
  \dot S_{\mathrm{env}}(t)=\sum_{\nu,k}\Delta s_{\nu k}(t)a_{\nu k}(t).
\end{equation}
Including the information contribution gives the entropy production rate
\begin{equation}\label{eq:feedback-entropy-production}
  \sigma_t:=\dot S_{\mathrm{sys}}(t)+\dot S_{\mathrm{env}}(t)-\dot I(t).
\end{equation}
We now relate $\sigma_t$ to $\sigma_t^{(0)}$, both evaluated on the same state $\rho_t$. 
Using Eq.~\eqref{eq:split-dynamics}, we find
\begin{equation}\label{eq:entropy-production-difference}
  \sigma_t-\sigma_t^{(0)}
  =-\Tr\!\left(\cB_t[\rho_t]\ln\rho_t\right)-\dot I(t)
  =\sum_{\nu,k}a_{\nu k}(t)
  \left[S(\cF_{\nu k}[\rho'_{\nu k,t}])-S(\rho'_{\nu k,t})\right],
\end{equation}
where the last equality follows from Eq.~\eqref{eq:information-entropy-form}. 
To determine the sign of each term, let $d$ be the system dimension. 
Contractivity of quantum relative entropy under a completely positive, trace-preserving map, together with unitality, implies
\begin{equation}\label{eq:unital-contractivity}
  D\!\left(\cF_{\nu k}[\rho']\middle\Vert\frac{\mathbb I}{d}\right)
  =D\!\left(\cF_{\nu k}[\rho']\middle\Vert\cF_{\nu k}\!\left[\frac{\mathbb I}{d}\right]\right)
  \leq D\!\left(\rho'\middle\Vert\frac{\mathbb I}{d}\right).
\end{equation}
Since $D(\rho'\Vert\mathbb I/d)=\ln d-S(\rho')$, this gives $S(\cF_{\nu k}[\rho'])\geq S(\rho')$. 
Every term on the right-hand side of Eq.~\eqref{eq:entropy-production-difference} is therefore nonnegative, and
\begin{equation}\label{eq:rate-comparison}
  \sigma_t\geq\sigma_t^{(0)}\geq0.
\end{equation}
For a process of duration $\tau$, the integrated entropy production is
\begin{equation}\label{eq:integrated-entropy-production}
  \Sigma_\tau:=\int_0^\tau\sigma_t\,\dd t=\Delta S_{\mathrm{sys}}+\Delta S_{\mathrm{env}}-I.
\end{equation}
Here, $\Sigma_\tau$ characterizes the entropy production associated with the system and its reservoirs, accounting for the information used by feedback.
Integration of Eq.~\eqref{eq:entropy-production-difference} yields
\begin{equation}\label{eq:integrated-difference}
  \Sigma_\tau-\Sigma_\tau^{(0)}=\int_0^\tau\dd t\,\sum_{\nu,k}a_{\nu k}(t)\left[S(\cF_{\nu k}[\rho'_{\nu k,t}])-S(\rho'_{\nu k,t})\right]\geq0,
\end{equation}
and hence
\begin{equation}\label{eq:integrated-comparison}
  \Sigma_\tau\geq\Sigma_\tau^{(0)}\geq0.
\end{equation}
This proves the generalized second law used in the main text.

\subsection{Proof of the bound}
\label{sec:bare-current-dissipation}

We now use the entropy-production expressions above to bound a weighted current.
For real, time-dependent weights satisfying $c_{\nu k^*}^{yx}(t)=-c_{\nu k}^{xy}(t)$, define the weighted current and the corresponding weighted dynamical state mobility as in the main text,
\begin{align}
  J_t&:=\sum_{\nu,k,x,y}c_{\nu k}^{xy}(t)w_{\nu k}^{xy}(t)p_y(t)=\frac12\sum_{\nu,k,x,y}c_{\nu k}^{xy}(t)j_{\nu k}^{xy}(t),
  \label{eq:weighted-current}\\
  m_t&:=\frac12\sum_{\nu,k,x,y}[c_{\nu k}^{xy}(t)]^2\ell_{\nu k}^{xy}(t).
  \label{eq:kinetic-coefficient}
\end{align}
The second expression for $J_t$ follows by relabeling the reverse transitions and using the antisymmetry of $c$. 
Applying the Cauchy--Schwarz inequality to the transition sum yields
\begin{equation}\label{eq:bare-instantaneous-current}
  \begin{split}
    [J_t]^2&=\left[\frac12\sum_{\nu,k,x,y}\left(c_{\nu k}^{xy}\sqrt{\ell_{\nu k}^{xy}}\right)\frac{j_{\nu k}^{xy}}{\sqrt{\ell_{\nu k}^{xy}}}\right]^2\\
    &\leq\left[\frac12\sum_{\nu,k,x,y}(c_{\nu k}^{xy})^2\ell_{\nu k}^{xy}\right]\left[\frac12\sum_{\nu,k,x,y}\frac{(j_{\nu k}^{xy})^2}{\ell_{\nu k}^{xy}}\right]=m_t\sigma_t^{(0)}.
  \end{split}
\end{equation}
Let $M_\tau:=\int_0^\tau m_t\,\dd t$. A second application of the Cauchy--Schwarz inequality, now in time, gives
\begin{equation}\label{eq:bare-finite-time-current}
  \begin{split}
    \left[\int_0^\tau J_t\,\dd t\right]^2&\leq\left[\int_0^\tau|J_t|\,\dd t\right]^2\leq\left[\int_0^\tau\sqrt{m_t\sigma_t^{(0)}}\,\dd t\right]^2\\
    &\leq\left[\int_0^\tau m_t\,\dd t\right]\left[\int_0^\tau\sigma_t^{(0)}\,\dd t\right]=M_\tau\Sigma_\tau^{(0)}.
  \end{split}
\end{equation}
Combining the bare current bound in Eq.~\eqref{eq:bare-finite-time-current} with the entropy-production comparison in Eq.~\eqref{eq:integrated-comparison}, we obtain
\begin{equation}\label{eq:feedback-finite-time-current}
  \left[\int_0^\tau |J_t|\,\dd t\right]^2\leq M_\tau\Sigma_\tau^{(0)}\leq M_\tau\Sigma_\tau. 
\end{equation}
This proves the finite-time current--dissipation relation in Eq.~(5) of the main text.

\section{Applications}
\label{sec:applications}

\subsection{Quantum speed limit}
\label{sec:quantum-speed-limit}

We now apply the current--dissipation relation to bound the time required for a state transformation. 
For this proof, we label the instantaneous eigenvalues in decreasing order, $p_y(t)=p_y^\downarrow(t)$. 
The trace-like distance between the initial and final states is defined as
\begin{equation}\label{eq:spectral-distance}
  d_T(\rho_0,\rho_\tau):=\sum_y|p_y(\tau)-p_y(0)|.
\end{equation}
% For piecewise continuously differentiable dynamics, the ordered
% eigenvalues are absolutely continuous and satisfy
% $\dot p_y(t)=\langle y_t|\dot\rho_t|y_t\rangle$ almost everywhere.
% Within a degenerate eigenspace, the basis may be chosen to diagonalize
% the restriction of $\dot\rho_t$ to that subspace. Thus eigenvalue
% crossings do not affect the integrated relations below.
Define the time-independent signs and the corresponding Hermitian operator
\begin{equation}\label{eq:spectral-operator}
  z_x:=\operatorname{sgn}[p_x(\tau)-p_x(0)],
  \qquad
  Z_t:=\sum_y z_y\proj{y_t},
\end{equation}
where $\operatorname{sgn}(0)=0$. 
By construction, $[Z_t,\rho_t]=0$ and $\|Z_t\|_\infty\leq1$, where $\|\cdot\|_\infty$ is the operator norm. 
Integrating the eigenvalue derivatives gives
\begin{equation}\label{eq:spectral-distance-integral}
  d_T(\rho_0,\rho_\tau)=\sum_y z_y[p_y(\tau)-p_y(0)]=\int_0^\tau\sum_y z_y\dot p_y(t)\,\dd t=\int_0^\tau\Tr(Z_t\dot\rho_t)\,\dd t.
\end{equation}
Using the decomposition in Eq.~\eqref{eq:split-dynamics}, we separate the bare contribution and the direct action of feedback,
\begin{equation}\label{eq:spectral-distance-split}
  d_T(\rho_0,\rho_\tau)=\int_0^\tau\Tr\!\left(Z_t\cL_t^0[\rho_t]\right)\dd t+\int_0^\tau\Tr\!\left(Z_t\cB_t[\rho_t]\right)\dd t.
\end{equation}

To express the first contribution as a weighted current, choose
\begin{equation}\label{eq:spectral-weights}
  c_{\nu k}^{xy}=z_x-z_y.
\end{equation}
These weights satisfy $c_{\nu k^*}^{yx}=-c_{\nu k}^{xy}$. 
The Hamiltonian contribution vanishes because $\Tr(Z_t[H_t,\rho_t])=\Tr(H_t[\rho_t,Z_t])=0$. 
For each dissipative term, expansion in the instantaneous eigenbasis gives
\begin{equation}\label{eq:dissipative-spectral-current}
  \Tr\!\left(Z_t\Diss{L_{\nu k}}[\rho_t]\right)=\sum_{x,y}z_x w_{\nu k}^{xy}p_y-\sum_{x,y}z_y w_{\nu k}^{xy}p_y=\sum_{x,y}(z_x-z_y)w_{\nu k}^{xy}p_y.
\end{equation}
Summing over all jump channels therefore yields
\begin{equation}\label{eq:spectral-current-coefficient}
  \Tr\!\left(Z_t\cL_t^0[\rho_t]\right)=\sum_{\nu,k,x,y}(z_x-z_y)w_{\nu k}^{xy}p_y=J_t,
  \qquad
  m_t=\frac12\sum_{\nu,k,x,y}(z_x-z_y)^2\ell_{\nu k}^{xy}(t).
\end{equation}

For the feedback contribution, we now introduce
\begin{equation}\label{eq:drift-norm}
  b_t:=\|\cB_t[\rho_t]\|_1=\left\|\sum_{\nu,k}a_{\nu k}(t)\left(\cF_{\nu k}[\rho'_{\nu k,t}]-\rho'_{\nu k,t}\right)\right\|_1,
  \qquad 
  \|Y\|_1:=\Tr\sqrt{Y^{\dagger}Y}.
\end{equation}
H\"older's inequality and $\|Z_t\|_\infty\leq1$ imply
\begin{equation}\label{eq:feedback-spectral-speed}
  \left|\Tr\!\left(Z_t\cB_t[\rho_t]\right)\right|\leq\|Z_t\|_\infty\|\cB_t[\rho_t]\|_1\leq b_t.
\end{equation}
Substituting Eqs.~\eqref{eq:spectral-current-coefficient} and \eqref{eq:feedback-spectral-speed} into Eq.~\eqref{eq:spectral-distance-split}, and applying Eq.~\eqref{eq:feedback-finite-time-current}, gives
\begin{equation}\label{eq:integrated-spectral-bound}
  \begin{split}
    d_T(\rho_0,\rho_\tau)&\leq\int_0^\tau |J_t|\,\dd t+\int_0^\tau b_t\,\dd t\\
    &\leq\sqrt{M_\tau\Sigma_\tau}+\int_0^\tau b_t\,\dd t.
  \end{split}
\end{equation}
Define the time averages
\begin{equation}\label{eq:speed-limit-averages}
  \overline{m}:=\frac{M_\tau}{\tau},
  \qquad
  \overline\sigma:=\frac{\Sigma_\tau}{\tau},
  \qquad
  \overline b:=\frac1\tau\int_0^\tau b_t\,\dd t.
\end{equation}
Equation~\eqref{eq:integrated-spectral-bound} becomes $d_T(\rho_0,\rho_\tau)\leq\tau(\sqrt{\overline{m}\,\overline\sigma}+\overline b)$, which yields
\begin{equation}\label{eq:feedback-quantum-speed-limit}
  \tau\geq\frac{d_T(\rho_0,\rho_\tau)}{\sqrt{\overline{m}\,\overline\sigma}+\overline b}.
\end{equation}
This is the quantum speed limit in Eq.~(6) of the main text. 
The quotient applies when the denominator is positive; if it vanishes, Eq.~\eqref{eq:integrated-spectral-bound} requires $d_T=0$. 
Feedback enters through both the information contribution to $\Sigma_\tau$ and the direct drift $b_t$.

For completeness, the logarithmic-mean inequality $\ell_{\nu k}^{xy}\leq(w_{\nu k}^{xy}p_y+w_{\nu k^*}^{yx}p_x)/2$ and $|z_x-z_y|\leq2$ give
\begin{equation}\label{eq:spectral-coefficient-activity}
  m_t\leq2\sum_{\nu,k,x,y}\ell_{\nu k}^{xy}(t)\leq\sum_{\nu,k,x,y}\left(w_{\nu k}^{xy}p_y+w_{\nu k^*}^{yx}p_x\right)=2a_t,
\end{equation}
where the last equality follows by relabeling the reverse transitions. 
Thus Eq.~\eqref{eq:feedback-quantum-speed-limit} is at least as strong as the corresponding activity bound,
\begin{equation}\label{eq:activity-quantum-speed-limit}
  \tau\geq
  \frac{d_T(\rho_0,\rho_\tau)}
  {\sqrt{2\overline a\,\overline\sigma}+\overline b},
  \qquad
  \overline a:=\frac1\tau\int_0^\tau a_t\,\dd t.
\end{equation}
In the absence of feedback, $I=0$, $\overline b=0$, and $\Sigma_\tau=\Sigma_\tau^{(0)}$. 
The comparison with the activity bound therefore also holds without feedback.

\subsection{Steady state power--efficiency trade-off}
\label{sec:stationary-engine-bound}

\subsubsection{Isothermal information engines}

We next consider a stationary information engine operating in an isothermal environment at inverse temperature $\beta$. 
All quantities in this subsection are evaluated in the stationary state $\rho$. 
As in the main text, the subscript $\mathrm{ss}$ denotes stationary values of the weighted current, mobility, entropy production rate, and weighted dynamical activity. 
We consider engines whose useful output power can be expressed as a weighted current of the monitored jumps. 
For a given output power $P$, we choose antisymmetric weights satisfying
\begin{equation}\label{eq:engine-output-current}
  c_{\nu k^*}^{yx}=-c_{\nu k}^{xy},
  \qquad
  P=J_{\mathrm{ss}}=\sum_{\nu,k,x,y}c_{\nu k}^{xy}w_{\nu k}^{xy}p_y.
\end{equation}
Different choices satisfying this condition can yield different weighted mobilities.
A concrete example is a particle engine with chemical output power $P=\Delta\mu\,J_N$, where $\Delta\mu$ is the chemical potential difference and $J_N$ is the net particle current. 

As in the main text, let $\dot Q=\beta^{-1}\sum_{\nu,k}\Delta s_{\nu k}a_{\nu k}$ be the total heat current released into the reservoirs and $P_F$ the net power supplied to the system by the feedback controller. 
The stationary first law and the reservoir entropy balance read
\begin{equation}\label{eq:engine-first-law}
  P=-\dot Q+P_F,
  \qquad
  \dot S_{\mathrm{env}}=\beta\dot Q,
\end{equation}
where
\begin{equation}\label{eq:engine-feedback-power}
  P_F=\Tr\!\left(H\cB[\rho]\right)=\sum_{\nu,k}a_{\nu k}\Tr\!\left\{H\left(\cF_{\nu k}[\rho'_{\nu k}]-\rho'_{\nu k}\right)\right\}.
\end{equation}

To illustrate this energy balance, consider a particle engine with a time-independent Hamiltonian $H$ and particle-number-conserving feedback.
Let $N$ denote the particle-number operator, with $[H,N]=0$.
For energy- and particle-resolved jumps, $[H,L_{\nu k}]=\epsilon_{\nu k}L_{\nu k}$ and $[N,L_{\nu k}]=n_{\nu k}L_{\nu k}$, where $\epsilon_{\nu k}$ and $n_{\nu k}$ are the energy and particle number transferred into the system by a bare jump.
For a reservoir with chemical potential $\mu_\nu$, the entropy change is $\Delta s_{\nu k}=-\beta(\epsilon_{\nu k}-\mu_\nu n_{\nu k})$.
Using $E=\Tr(H\rho)$ and $\Tr(H\Diss{L_{\nu k}}[\rho])=\epsilon_{\nu k}a_{\nu k}$, we obtain the stationary energy balance, heat current, and chemical output power as
\begin{equation}\label{eq:particle-engine-energy-balance}
  \begin{aligned}
    0=\dot E&=\sum_{\nu,k}\epsilon_{\nu k}a_{\nu k}+P_F,\\
    \dot Q&=-\sum_{\nu,k}(\epsilon_{\nu k}-\mu_\nu n_{\nu k})a_{\nu k},\\
    P&=-\sum_{\nu,k}\mu_\nu n_{\nu k}a_{\nu k}.
  \end{aligned}
\end{equation}
Combining these expressions gives $P=-\dot Q+P_F$, as in Eq.~\eqref{eq:engine-first-law}.
The choice $c_{\nu k}^{xy}=-\mu_\nu n_{\nu k}$ is one admissible representation of this output power in Eq.~\eqref{eq:engine-output-current}.

Both the energy supplied to the system by the controller and the information consumed by feedback enter the input resource in the efficiency,
\begin{equation}\label{eq:engine-information-efficiency}
  \eta_I:=\frac{P}{P_F-\dot I/\beta}.
\end{equation}
We consider the engine regime $P>0$. Since the system entropy is constant in the stationary state, Eqs.~\eqref{eq:feedback-entropy-production} and \eqref{eq:engine-first-law} give
\begin{equation}\label{eq:engine-entropy-efficiency}
  \begin{split}
    \sigma_{\mathrm{ss}}
    &=\beta\dot Q-\dot I
    =\beta\left(P_F-\frac{\dot I}{\beta}-P\right)\\
    &=\beta P\frac{1-\eta_I}{\eta_I}\geq0.
  \end{split}
\end{equation}
In particular, $P_F-\dot I/\beta\geq P>0$, so the efficiency is well defined and satisfies $0<\eta_I\leq1$.

For an arbitrary observation time $\tau$, stationarity gives $\int_0^\tau |J_t|\,\dd t=\tau P$, $M_\tau=\tau m_{\mathrm{ss}}$, and $\Sigma_\tau=\tau\sigma_{\mathrm{ss}}$. 
Thus Eq.~\eqref{eq:feedback-finite-time-current} reduces to
\begin{equation}\label{eq:stationary-power-current-bound}
  P^2\leq m_{\mathrm{ss}}\sigma_{\mathrm{ss}}^{(0)}\leq m_{\mathrm{ss}}\sigma_{\mathrm{ss}},
  \qquad
  m_{\mathrm{ss}}=\frac12\sum_{\nu,k,x,y}(c_{\nu k}^{xy})^2\ell_{\nu k}^{xy}.
\end{equation}
Substituting Eq.~\eqref{eq:engine-entropy-efficiency} and dividing by $P>0$ yields
\begin{equation}\label{eq:stationary-power-efficiency-bound}
  P\leq\beta m_{\mathrm{ss}}\frac{1-\eta_I}{\eta_I},
\end{equation}
which is Eq.~(12) of the main text. 
The weighted mobility $m_{\mathrm{ss}}$ is determined by the stationary state and the weights representing the output power. 
For bounded $m_{\mathrm{ss}}$, the output power vanishes as $\eta_I\to1$.

The logarithmic mean inequality also gives an activity form of the bound. 
Define the weighted stationary dynamical activity
\begin{equation}\label{eq:engine-weighted-activity}
  a_{\mathrm{ss}}:=\sum_{\nu,k,x,y}(c_{\nu k}^{xy})^2w_{\nu k}^{xy}p_y.
\end{equation}
Using $(c_{\nu k^*}^{yx})^2=(c_{\nu k}^{xy})^2$ and relabeling reverse transitions, we obtain
\begin{equation}\label{eq:engine-coefficient-activity}
  m_{\mathrm{ss}}\leq\frac14\sum_{\nu,k,x,y}(c_{\nu k}^{xy})^2\left(w_{\nu k}^{xy}p_y+w_{\nu k^*}^{yx}p_x\right)=\frac12a_{\mathrm{ss}}.
\end{equation}
Consequently,
\begin{equation}\label{eq:stationary-power-activity-bound}
  P\leq\frac{\beta a_{\mathrm{ss}}}{2}\frac{1-\eta_I}{\eta_I}.
\end{equation}
When the weights depend only on the monitored channel, $a_{\mathrm{ss}}=\sum_{\nu,k}c_{\nu k}^2a_{\nu k}$. 
The bound in Eq.~\eqref{eq:stationary-power-efficiency-bound} retains the logarithmic means and is therefore at least as strong as Eq.~\eqref{eq:stationary-power-activity-bound}.

\subsubsection{Extension to reservoirs at different temperatures}

The same argument applies to stationary engines coupled to reservoirs
with different inverse temperatures $\beta_\nu$, provided the output
power has the current representation in Eq.~\eqref{eq:engine-output-current}. 
With heat currents positive into the reservoirs, the stationary first and second laws are
\begin{equation}\label{eq:multibath-engine-balances}
  P=-\sum_\nu\dot Q_\nu+P_F,
  \qquad
  \sigma_{\mathrm{ss}}=\sum_\nu\beta_\nu\dot Q_\nu-\dot I\geq0.
\end{equation}
Let $\beta_*:=\max_\nu\beta_\nu$ be the inverse temperature of the
coldest reservoir. 
Relative to this reservoir, define the input resource rate and the corresponding efficiency by
\begin{equation}\label{eq:multibath-resource-efficiency}
  R:=P_F-\frac{\dot I}{\beta_*}-\sum_\nu\left(1-\frac{\beta_\nu}{\beta_*}\right)\dot Q_\nu,
  \qquad
  \eta_{\mathrm{ex}}:=\frac{P}{R}.
\end{equation}
The last term in $R$ accounts for the part of the heat input available for work relative to the coldest reservoir. 
Eliminating the heat currents between Eqs.~\eqref{eq:multibath-engine-balances} and \eqref{eq:multibath-resource-efficiency} gives
\begin{equation}\label{eq:multibath-entropy-efficiency}
  \sigma_{\mathrm{ss}}=\beta_*(R-P)=\beta_*P\frac{1-\eta_{\mathrm{ex}}}{\eta_{\mathrm{ex}}}.
\end{equation}
For $P>0$, nonnegative entropy production implies $R\geq P>0$ and $0<\eta_{\mathrm{ex}}\leq1$. 
Combining Eq.~\eqref{eq:multibath-entropy-efficiency} with Eq.~\eqref{eq:stationary-power-current-bound} yields
\begin{equation}\label{eq:multibath-power-efficiency-bound}
  P\leq\beta_* m_{\mathrm{ss}}\frac{1-\eta_{\mathrm{ex}}}{\eta_{\mathrm{ex}}}.
\end{equation}
When all reservoirs have the same temperature, $\eta_{\mathrm{ex}}=\eta_I$ and this expression reduces to Eq.~\eqref{eq:stationary-power-efficiency-bound}.

% For a two-reservoir heat engine without feedback, $P_F=\dot I=0$ and
% $\beta_h<\beta_c=\beta_*$. Defining the thermal efficiency
% $\eta:=-P/\dot Q_h$ and the Carnot efficiency
% $\eta_C:=1-\beta_h/\beta_c$, we have
% $R=-\eta_C\dot Q_h$ and $\eta_{\mathrm{ex}}=\eta/\eta_C$.
% Equation~\eqref{eq:multibath-power-efficiency-bound} then becomes
% \begin{equation}
%   P\leq\beta_c m_{\mathrm{ss}}\frac{\eta_C-\eta}{\eta},
%   \label{eq:two-bath-power-efficiency-bound}
% \end{equation}
% where $m_{\mathrm{ss}}$ remains the weighted mobility associated with the output-power
% weights in Eq.~\eqref{eq:engine-output-current}.

\subsection{Information heat pump}
\label{sec:heat-pump-bound}

We next consider a stationary heat pump coupled to two thermal reservoirs with inverse temperatures $0<\beta_h<\beta_c$. 
The system Hamiltonian $H$ is time independent, and the controller supplies energy through the feedback operations. 
We retain the convention that heat currents are positive into the reservoirs. 
On the heat-pump regime, the system absorbs heat from the cold reservoir and releases it to the hot reservoir, so that $\dot Q_c<0$ and $\dot Q_h>0$, and the heating power is
\begin{equation}\label{eq:heat-pump-heating-power}
  P_{\mathrm{hp}}:=\dot Q_h>0.
\end{equation}
The stationary condition gives
\begin{equation}\label{eq:heat-pump-first-law}
  \dot Q_c+\dot Q_h=P_F,
  \qquad
  \dot Q_c=P_F-P_{\mathrm{hp}},
\end{equation}
where $P_F$ is the feedback energy input defined in Eq.~\eqref{eq:engine-feedback-power}. 
Including the information contribution, we define the heating coefficient of performance (COP) by
\begin{equation}\label{eq:heat-pump-cop}
  \kappa:=\frac{P_{\mathrm{hp}}}{P_F-\dot I/\beta_c}.
\end{equation}
As in the information engine efficiency, the denominator accounts for both controller energy and information consumption.

Since the system entropy is constant in the stationary state, the entropy production rate reads
\begin{equation}\label{eq:heat-pump-entropy-cop}
  \begin{split}
    \sigma_{\mathrm{ss}}&=\beta_c\dot Q_c+\beta_h\dot Q_h-\dot I\\
    &=\beta_c\left(P_F-\frac{\dot I}{\beta_c}\right)-(\beta_c-\beta_h)P_{\mathrm{hp}}=\beta_cP_{\mathrm{hp}}\left(\frac1\kappa-\eta_C\right),
  \end{split}
\end{equation}
where $\eta_C:=1-\beta_h/\beta_c>0$ is the Carnot efficiency. 
Nonnegative entropy production implies $P_F-\dot I/\beta_c\geq\eta_C P_{\mathrm{hp}}>0$, so $\kappa$ is well defined on the heat-pump regime and satisfies
\begin{equation}\label{eq:heat-pump-carnot-cop}
  0<\kappa\leq\kappa_C,
  \qquad
  \kappa_C:=\frac1{\eta_C}=\frac{\beta_c}{\beta_c-\beta_h}.
\end{equation}
Here, $\kappa_C$ is the Carnot heating COP. 
We now strengthen this inequality by retaining the dissipation required by both heat currents.

\subsubsection{Heat current bounds for the two reservoirs}

For a thermal reservoir, the heat released into the reservoir in a bare jump $(\nu,k)$ is $\Delta s_{\nu k}/\beta_\nu$. 
We assume
\begin{equation}\label{eq:heat-pump-thermal-jumps}
  [H,L_{\nu k}]=-\frac{\Delta s_{\nu k}}{\beta_\nu}L_{\nu k},
  \qquad
  \Delta s_{\nu k^*}=-\Delta s_{\nu k},
  \quad \nu\in\{c,h\}.
\end{equation}
The energy exchanged during the subsequent feedback operation is included in $P_F$. 
The commutator relation above gives
\begin{equation}\label{eq:thermal-jump-energy-current}
  \Tr\!\left(H\Diss{L_{\nu k}}[\rho]\right)=\Tr\!\left[\left(L_{\nu k}^{\dagger}HL_{\nu k}-\frac12\{L_{\nu k}^{\dagger}L_{\nu k},H\}\right)\rho\right]=-\frac{\Delta s_{\nu k}}{\beta_\nu}a_{\nu k}.
\end{equation}
Thus the heat current into reservoir $\nu$ has the equivalent forms
\begin{equation}\label{eq:heat-pump-bare-heat-current}
  \dot Q_\nu=-\sum_k\Tr\!\left(H\Diss{L_{\nu k}}[\rho]\right)=\frac1{\beta_\nu}\sum_k\Delta s_{\nu k}a_{\nu k}=\frac1{2\beta_\nu}\sum_{k,x,y}\Delta s_{\nu k}j_{\nu k}^{xy}.
\end{equation}
The last equality follows by pairing each transition with its reverse and using $\Delta s_{\nu k^*}=-\Delta s_{\nu k}$. 
For each reservoir, define the weighted mobility associated with its heat current and the bare entropy production rate by
\begin{equation}\label{eq:heat-pump-bath-coefficients}
  m_{Q_\nu}:=\frac1{2\beta_\nu^2}\sum_{k,x,y}(\Delta s_{\nu k})^2\ell_{\nu k}^{xy},
  \qquad
  \sigma_{\nu,\mathrm{ss}}^{(0)}:=\frac12\sum_{k,x,y}
  \frac{(j_{\nu k}^{xy})^2}{\ell_{\nu k}^{xy}},
  \qquad
  \sigma_{\mathrm{ss}}^{(0)}=\sigma_{c,\mathrm{ss}}^{(0)}+\sigma_{h,\mathrm{ss}}^{(0)}.
\end{equation}
The mobility $m_{Q_\nu}$ is obtained by assigning the energy weights $\Delta s_{\nu k}/\beta_\nu$ to the jumps induced by reservoir $\nu$. 
Applying the current--dissipation relation in Eq.~\eqref{eq:bare-instantaneous-current} separately to the jumps induced by each reservoir gives
\begin{equation}\label{eq:heat-pump-partial-current-bound}
  \dot Q_\nu^2\leq m_{Q_\nu}\sigma_{\nu,\mathrm{ss}}^{(0)}.
\end{equation}
Dividing by $m_{Q_\nu}$, summing over the two reservoirs, and using $\sigma_{\mathrm{ss}}^{(0)}\leq\sigma_{\mathrm{ss}}$, we obtain
\begin{equation}\label{eq:heat-pump-two-current-bound}
  \frac{\dot Q_c^2}{m_{Q_c}}+\frac{\dot Q_h^2}{m_{Q_h}}\leq\sigma_{c,\mathrm{ss}}^{(0)}+\sigma_{h,\mathrm{ss}}^{(0)}=\sigma_{\mathrm{ss}}^{(0)}\leq\sigma_{\mathrm{ss}}.
\end{equation}

\subsubsection{Power--COP trade-off}

Substituting Eqs.~\eqref{eq:heat-pump-first-law} and \eqref{eq:heat-pump-entropy-cop} into Eq.~\eqref{eq:heat-pump-two-current-bound} yields
\begin{equation}\label{eq:heat-pump-general-bound}
  \frac{(P_{\mathrm{hp}}-P_F)^2}{m_{Q_c}}+\frac{P_{\mathrm{hp}}^2}{m_{Q_h}}\leq\beta_cP_{\mathrm{hp}}\left(\frac1\kappa-\frac1{\kappa_C}\right).
\end{equation}
This is the heating-power--COP trade-off including the energy supplied by feedback. 
The first term on the left-hand side accounts for the heat extracted from the cold reservoir, while the second accounts for the heat delivered to the hot reservoir.

When the feedback supplies no net energy, $P_F=0$, the two heat currents have equal magnitude. 
Defining the effective mobility
\begin{equation}\label{eq:heat-pump-effective-coefficient}
  m_{\mathrm{hp}}:=\left(\frac1{m_{Q_c}}+\frac1{m_{Q_h}}\right)^{-1}=\frac{m_{Q_c}m_{Q_h}}{m_{Q_c}+m_{Q_h}},
\end{equation}
Eq.~\eqref{eq:heat-pump-general-bound} becomes $P_{\mathrm{hp}}^2/m_{\mathrm{hp}}\leq\beta_cP_{\mathrm{hp}}(\kappa^{-1}-\kappa_C^{-1})$. 
Since $P_{\mathrm{hp}}>0$, we obtain
\begin{equation}\label{eq:heat-pump-energy-neutral-bound}
  P_{\mathrm{hp}}\leq\beta_c m_{\mathrm{hp}}\left(\frac1\kappa-\frac1{\kappa_C}\right).
\end{equation}
In this case, $\kappa=P_{\mathrm{hp}}/(-\dot I/\beta_c)$, and the heat pump is driven by information consumption. 
For bounded $m_{\mathrm{hp}}$, the heating power vanishes as $\kappa\to\kappa_C$.

\subsection{Cyclic heat engine}
\label{sec:two-reservoir-engine}

We now consider a cyclic heat engine operating between reservoirs at inverse temperatures $0<\beta_h<\beta_c$. 
We first give the derivation for a cycle of duration $\tau$, with $\rho_\tau=\rho_0$ and $H_\tau=H_0$. 
The thermal jump conditions in Eq.~\eqref{eq:heat-pump-thermal-jumps} hold at each instant along the protocol. 
Define the heat absorbed from the hot reservoir and the heat released to the cold reservoir by
\begin{equation}\label{eq:two-reservoir-cycle-heats}
  Q_h:=-\int_0^\tau\dot Q_h(t)\,\dd t>0,
  \qquad
  Q_c:=\int_0^\tau\dot Q_c(t)\,\dd t>0.
\end{equation}
Thus $Q_h$ and $Q_c$ are positive heat magnitudes, while the instantaneous currents $\dot Q_\nu(t)$ are positive into the reservoirs. 
Let $W>0$ be the useful work output and $E_F:=\int_0^\tau P_F(t)\,\mathrm{d}t$ the net energy supplied by the feedback controller. 
Periodicity of the system energy gives
\begin{equation}\label{eq:two-reservoir-cycle-first-law}
  W=Q_h-Q_c+E_F.
\end{equation}
The output power and thermal efficiency are
\begin{equation}\label{eq:two-reservoir-thermal-efficiency}
  P:=\frac{W}{\tau},
  \qquad
  \eta:=\frac{W}{Q_h}.
\end{equation}
We use the thermal efficiency, which does not include information consumption in the input resource, to facilitate comparison with existing power--efficiency bounds for heat engines without feedback.
An efficiency that includes the information contribution can also be defined, but we do not pursue this alternative here.
The Carnot efficiency is $\eta_C:=1-\beta_h/\beta_c$.

\subsubsection{Energy and information contributions}

Since the system entropy returns to its initial value after one cycle,
\begin{equation}\label{eq:two-reservoir-cycle-second-law}
  \Sigma_\tau=\beta_cQ_c-\beta_hQ_h-I,
  \qquad
  I=\int_0^\tau\dot I(t)\,\dd t.
\end{equation}
Define the feedback inputs relative to the absorbed heat,
\begin{equation}\label{eq:two-reservoir-resource-ratios}
  \zeta_F:=\frac{E_F}{Q_h},
  \qquad
  \zeta_I:=-\frac{I}{\beta_cQ_h},
  \qquad
  \zeta:=\zeta_F+\zeta_I.
\end{equation}
Information consumption corresponds to $\zeta_I>0$, while $\zeta_F$ is positive when the controller supplies energy and negative when it extracts energy. 
Dividing Eq.~\eqref{eq:two-reservoir-cycle-first-law} by $Q_h$ gives
\begin{equation}\label{eq:two-reservoir-heat-ratios}
  \frac{Q_c}{Q_h}=1-\eta+\zeta_F,
  \qquad
  \frac{Q_h+Q_c}{Q_h}=2-\eta+\zeta_F.
\end{equation}
Substitution into Eq.~\eqref{eq:two-reservoir-cycle-second-law} yields
\begin{equation}\label{eq:two-reservoir-efficiency-gap}
  \begin{split}
    \Sigma_\tau&=\beta_cQ_h\left(1-\eta+\zeta_F-\frac{\beta_h}{\beta_c}+\zeta_I\right)\\
    &=\beta_cQ_h(\eta_C+\zeta-\eta).
  \end{split}
\end{equation}
The engine regime therefore requires
\begin{equation}\label{eq:two-reservoir-engine-domain}
  \eta>0,
  \qquad 
  1-\eta+\zeta_F>0,
  \qquad 
  \eta\leq\eta_C+\zeta.
\end{equation}
Both feedback energy and information enter the entropy balance, whereas only feedback energy enters the ratio $Q_c/Q_h$.

\subsubsection{Power--efficiency bound}

We choose the weights $c_{hk}^{xy}(t)=-\Delta s_{hk}(t)/\beta_h$ and $c_{ck}^{xy}(t)=\Delta s_{ck}(t)/\beta_c$, so that
\begin{equation}\label{eq:two-reservoir-combined-current}
  J_t=\dot Q_c(t)-\dot Q_h(t),
  \qquad
  \int_0^\tau J_t\,\dd t=Q_h+Q_c.
\end{equation}
For each reservoir $\nu\in\{h,c\}$, the weighted mobility associated with its heat current and its cycle average are
\begin{equation}\label{eq:two-reservoir-average-coefficients}
  m_{Q_\nu}(t):=\frac1{2\beta_\nu^2}\sum_{k,x,y}[\Delta s_{\nu k}(t)]^2\ell_{\nu k}^{xy}(t),
  \qquad
  \overline{m}_{Q_\nu}:=\frac1\tau\int_0^\tau m_{Q_\nu}(t)\,\dd t.
\end{equation}
Since the weights enter quadratically, $m_t=m_{Q_h}(t)+m_{Q_c}(t)$, with cycle average
\begin{equation*}
  \overline{m}:=\frac1\tau\int_0^\tau m_t\,\dd t =\overline{m}_{Q_h}+\overline{m}_{Q_c}.
\end{equation*}
Using $|\int_0^\tau J_t\,\dd t|\leq\int_0^\tau|J_t|\,\dd t$, Eq.~\eqref{eq:feedback-finite-time-current} gives
\begin{equation}\label{eq:two-reservoir-combined-heat-bound}
  (Q_h+Q_c)^2\leq\tau\overline{m}\Sigma_\tau.
\end{equation}
Using $Q_h+Q_c=(2-\eta+\zeta_F)Q_h$, $\Sigma_\tau=\beta_cQ_h(\eta_C+\zeta-\eta)$, and $P=\eta Q_h/\tau$, we obtain
\begin{equation}\label{eq:two-reservoir-combined-power-bound}
  P\leq\beta_c\overline{m}\frac{\eta(\eta_C+\zeta-\eta)}{(2-\eta+\zeta_F)^2}.
\end{equation}
In the absence of feedback, $E_F=I=0$, so $\zeta_F=\zeta_I=0$ and $0<\eta\leq\eta_C$. The bound reduces to
\begin{equation}\label{eq:two-reservoir-no-feedback-bound}
  P\leq\beta_c\overline{m}\frac{\eta(\eta_C-\eta)}{(2-\eta)^2}.
\end{equation}
For bounded $\overline{m}$, the power vanishes as $\eta\to\eta_C$.

An activity form follows from Eq.~\eqref{eq:engine-coefficient-activity}, applied at each time. 
For the weights chosen above, the weighted dynamical activity and its cycle average are
\begin{equation}\label{eq:two-reservoir-weighted-activity}
  a_c(t)=\sum_{\nu=h,c}\sum_k
  \left[\frac{\Delta s_{\nu k}(t)}{\beta_\nu}\right]^2a_{\nu k}(t),
  \qquad
  \overline a_c:=\frac1\tau\int_0^\tau a_c(t)\,\dd t.
\end{equation}
Here, each jump rate is weighted by the square of the heat released into the reservoir in that jump. 
Since $m_t\leq a_c(t)/2$, we have
\begin{equation}\label{eq:two-reservoir-coefficient-activity}
  \overline{m}\leq\frac{\overline a_c}{2}.
\end{equation}
Equation~\eqref{eq:two-reservoir-combined-power-bound} therefore implies
\begin{equation}\label{eq:two-reservoir-activity-bound}
  P\leq\frac{\beta_c\overline a_c}{2}\frac{\eta(\eta_C+\zeta-\eta)}{(2-\eta+\zeta_F)^2}.
\end{equation}
The bound in Eq.~\eqref{eq:two-reservoir-combined-power-bound} retains the logarithmic means and is at least as strong as this activity form.

\section{Examples}
\label{sec:examples}

\subsection{Finite-time information erasure}
\label{sec:memory-erasure}

We now derive the results for the example of finite-time information erasure in the main text. 
For a system coupled to a single reservoir at inverse temperature $\beta$, let $Q$ denote the heat dissipated into the reservoir. 
Then $\Delta S_{\mathrm{env}}=\beta Q$, and Eq.~\eqref{eq:integrated-entropy-production} gives
\begin{equation}\label{eq:erasure-heat-balance}
  \beta Q=-\Delta S_{\mathrm{sys}}+I+\Sigma_\tau.
\end{equation}
Combining this identity with Eq.~\eqref{eq:integrated-spectral-bound}, we obtain
\begin{equation}\label{eq:erasure-finite-time-Landauer}
  \beta Q\geq-\Delta S_{\mathrm{sys}}+I+\frac{[d_T(\rho_0,\rho_\tau)-\tau\overline b]_+^2}{\tau\overline{m}},
  \qquad 
  [x]_+:=\max\{x,0\}.
\end{equation}
This is the finite-time generalized Landauer bound in the main text. 
We evaluate each term for the qubit reset with feedback and then derive the optimal protocol.

\subsubsection{Feedback dynamics and thermodynamic quantities}

The Hamiltonian and state are
\begin{equation}\label{eq:erasure-H-state}
  H_t=\frac{\lambda_t}{2}(\proj{1}-\proj{0}),
  \qquad
  \rho_t=[1-p_t]\proj{0}+p_t\proj{1},
  \qquad 
  \lambda_t\geq0.
\end{equation}
Thus $|0\rangle$ and $|1\rangle$ denote the ground and excited states, respectively. 
The boundary conditions are
\begin{equation}\label{eq:erasure-boundaries}
  p_0=\frac12,\qquad p_\tau=\varepsilon,
  \qquad 
  0<\varepsilon<\frac12.
\end{equation}
Let $\Gamma>0$ denote the coupling coefficient. 
The thermal jump operators and feedback operations are
\begin{align}
  L_0&=\sqrt{\Gamma R(\beta\lambda_t)}\,|0\rangle\langle1|,&L_1&=\ee^{-\beta\lambda_t/2}\sqrt{\Gamma R(\beta\lambda_t)}\,|1\rangle\langle0|,
  \label{eq:erasure-jumps}\\
  \cF_0[X]&=X,&\cF_1[X]&=U_{10}XU_{10}^{\dagger},
  \qquad 
  U_{10}=|0\rangle\langle1|+|1\rangle\langle0|.
  \label{eq:erasure-feedback}
\end{align}
The response function satisfies $R(\beta\lambda_t)>0$. 
The reservoir entropy changes associated with the two bare jumps are $\Delta s_0=\beta\lambda_t$ and $\Delta s_1=-\beta\lambda_t$. 
Since $U_{10}L_1=\ee^{-\beta\lambda_t/2}\sqrt{\Gamma R(\beta\lambda_t)}\,\proj{0}$, the Wiseman equation becomes
\begin{equation}\label{eq:erasure-master}
  \dot\rho_t=-i[H_t,\rho_t]+\Diss{L_0}[\rho_t]+\Diss{U_{10}L_1}[\rho_t].
\end{equation}
For the diagonal state in Eq.~\eqref{eq:erasure-H-state}, both the commutator and the last dissipator vanish. 
The bare jump rates and the actual population velocity are therefore
\begin{equation}\label{eq:erasure-population}
  a_0=\Gamma R(\beta\lambda_t)p_t,
  \qquad 
  a_1=\Gamma R(\beta\lambda_t)\ee^{-\beta\lambda_t}(1-p_t),
  \qquad 
  \dot p_t=-a_0=-\Gamma p_tR(\beta\lambda_t).
\end{equation}
An upward jump still exchanges heat with the reservoir, even though the subsequent feedback cancels its population change.

Define the affinity\label{eq:erasure-affinity}
\begin{equation}
  \xi(p_t):=\beta\lambda_t-\ln\frac{1-p_t}{p_t}=\ln\frac{a_0}{a_1},
  \qquad 
  a_1=a_0\ee^{-\xi(p_t)}.
\end{equation}
The system entropy is $h_2(p):=-p\ln p-(1-p)\ln(1-p)$, hence
\begin{equation}\label{eq:erasure-system-entropy}
  \dot S_{\mathrm{sys}}=\dot p_t\,\ln\frac{1-p_t}{p_t}=-a_0\ln\frac{1-p_t}{p_t},
  \qquad
  \Delta S_{\mathrm{sys}}=h_2(\varepsilon)-\ln2.
\end{equation}
Only the feedback following an upward jump contributes to the information rate. 
It maps the conditional state $\proj{1}$ to $\proj{0}$. 
Using $D(\proj{0}\Vert\rho_t)=-\ln(1-p_t)$ and $D(\proj{1}\Vert\rho_t)=-\ln p_t$ in Eq.~\eqref{eq:information-rate}, we obtain
\begin{equation}\label{eq:erasure-information-rate}
  \dot I=a_1\left[D(\proj{0}\Vert\rho_t)-D(\proj{1}\Vert\rho_t)\right]=a_1\ln\frac{p_t}{1-p_t}\leq0.
\end{equation}
Each downward jump deposits energy $\lambda_t$ in the reservoir and each upward jump removes the same amount. 
Hence, we have
\begin{equation}\label{eq:erasure-heat-rate}
  \dot Q=\lambda_t(a_0-a_1),
  \qquad 
  \dot S_{\mathrm{env}}=\beta\lambda_t(a_0-a_1).
\end{equation}
Substituting Eqs.~\eqref{eq:erasure-system-entropy}--\eqref{eq:erasure-heat-rate} into Eq.~\eqref{eq:feedback-entropy-production} gives
\begin{equation}\label{eq:erasure-entropy-rate}
  \begin{split}
    \sigma_t&=\dot S_{\mathrm{sys}}+\beta\dot Q-\dot I=(a_0-a_1)\left[\beta\lambda_t-\ln\frac{1-p_t}{p_t}\right]\\
    &=a_0\xi(p_t)(1-\ee^{-\xi(p_t)})\geq0.
  \end{split}
\end{equation}
Changing the integration variable from $t$ to $p$ using $\dd t=-\dd p/[\Gamma pR(\beta\lambda)]$ yields
\begin{align}
  \Sigma[\lambda]&=\int_{\varepsilon}^{1/2}\xi(p)[1-\ee^{-\xi(p)}]\,\dd p,
  \label{eq:erasure-entropy-functional}\\
  \widehat\tau[\lambda]:=\Gamma\tau[\lambda]&=\int_{\varepsilon}^{1/2}\frac{\dd p}{pR[\beta\lambda(p)]},
  \label{eq:erasure-time-functional}\\
  \beta Q[\lambda]&=\int_{\varepsilon}^{1/2}\beta\lambda(p)[1-\ee^{-\xi(p)}]\,\dd p,
  \qquad
  I[\lambda]=-\int_{\varepsilon}^{1/2}\ee^{-\xi(p)}\ln\frac{1-p}{p}\,\dd p.
  \label{eq:erasure-heat-information-functionals}
\end{align}
These expressions reproduce Eq.~\eqref{eq:erasure-heat-balance}. 
Since $\Delta S_{\mathrm{sys}}$ is fixed by the boundary conditions, minimizing $\Sigma_\tau$ is equivalent to minimizing $\beta Q-I$.

For this model, the quantities entering the speed limit are
\begin{equation}\label{eq:erasure-speed-limit-quantities}
  d_T(\rho_0,\rho_\tau)=1-2\varepsilon=:D,
  \qquad 
  b_t=2a_1,
  \qquad 
  m_t=4a_0\frac{1-\ee^{-\xi(p_t)}}{\xi(p_t)}.
\end{equation}
The ratio $(1-\ee^{-\xi})/\xi$ is defined by continuity as $1$ at $\xi=0$. 
Their time integrals are
\begin{align}
  B_\tau:=\int_0^\tau b_t\,\mathrm{d}t
  &=2\int_{\varepsilon}^{1/2}\ee^{-\xi(p)}\,\mathrm{d}p,
  \label{eq:erasure-feedback-displacement}\\
  M_\tau=\int_0^\tau m_t\,\mathrm{d}t
  &=4\int_{\varepsilon}^{1/2}
  \frac{1-\ee^{-\xi(p)}}{\xi(p)}\,\mathrm{d}p.
  \label{eq:erasure-integrated-coefficient}
\end{align}
Equation~\eqref{eq:integrated-spectral-bound} therefore gives
\begin{equation}\label{eq:erasure-QSL}
  \Sigma_\tau\geq\Sigma_{\mathrm{QSL}}:=\frac{[D-B_\tau]_+^2}{M_\tau}.
\end{equation}
This bound is evaluated along the optimal protocols to obtain the dashed curves in the figure of the main text.

\subsubsection{Finite-time attainment of the generalized Landauer bound}

The integrand in Eq.~\eqref{eq:erasure-entropy-functional} is nonnegative and vanishes only at $\xi=0$. 
A protocol with $\Sigma_\tau=0$ therefore obeys
\begin{equation}\label{eq:erasure-reversible-protocol}
  \beta\lambda_{\mathrm{rev}}(p)=\ln\frac{1-p}{p},
  \qquad 
  a_0=a_1.
\end{equation}
Although the bare current vanishes, Eq.~\eqref{eq:erasure-population} still gives $\dot p_t=-\Gamma p_tR\!\left[\ln\frac{1-p_t}{p_t}\right]$. 
Its duration is
\begin{equation}\label{eq:erasure-reversible-time}
  \Gamma\tau^*=\int_{\varepsilon}^{1/2}\frac{\dd p}{pR\!\left[\ln\frac{1-p}{p}\right]}=\int_0^{\ln[(1-\varepsilon)/\varepsilon]}\frac{\ee^{\beta\lambda}}{(1+\ee^{\beta\lambda})R(\beta\lambda)}\,\dd (\beta\lambda).
\end{equation}
For a bosonic reservoir with a power-law spectral density, the response function is
\begin{equation}\label{eq:erasure-bosonic-response}
  R(\beta\lambda)=\frac{\lambda^s}{1-\ee^{-\beta\lambda}},
  \qquad 
  s\geq0.
\end{equation}
Equation~\eqref{eq:erasure-reversible-time} then becomes
\begin{equation}\label{eq:erasure-bosonic-reversible-time}
  \Gamma\tau^*=\beta^s\int_0^{\ln[(1-\varepsilon)/\varepsilon]}
  \frac{\tanh(\beta\lambda/2)}{(\beta\lambda)^s}\,\dd(\beta\lambda).
\end{equation}
The duration $\tau^*$ is finite precisely when $0\leq s<2$.

For the protocol in Eq.~\eqref{eq:erasure-reversible-protocol},
\begin{equation}\label{eq:erasure-reversible-balances}
  Q=0,\qquad 
  I=\Delta S_{\mathrm{sys}}=h_2(\varepsilon)-\ln2,
  \qquad 
  B_{\tau^*}=D,\qquad \Sigma_{\tau^*}=0.
\end{equation}
The entropy reduction is balanced by information consumption, and feedback supplies the full spectral displacement. 
The generalized Landauer bound and the speed limit are both saturated.

If $\tau^*$ is finite and $\tau\geq\tau^*$, the system can wait in its initial state for $\tau-\tau^*$ and then follow the protocol in Eq.~\eqref{eq:erasure-reversible-protocol}. 
With no heat flow or information consumption during the waiting interval, we obtain
\begin{equation}\label{eq:erasure-zero-plateau}
  \Sigma_{\min}(\tau,\varepsilon)=0,
  \qquad 
  \tau\geq\tau^*.
\end{equation}

\subsubsection{Variational derivation of the optimal protocol}

For a prescribed duration $\tau$, we minimize Eq.~\eqref{eq:erasure-entropy-functional} subject to $\widehat\tau[\lambda]\leq\Gamma\tau$, allowing for a waiting interval. 
The initial and final populations are fixed, while the endpoint gaps and the rate of change of the gap are unconstrained.

We consider protocols that complete erasure in time $\tau$ without a waiting interval.
By writing $\widehat t:=\Gamma t$ as a function of $p$, Eq.~\eqref{eq:erasure-population} gives
\begin{equation}\label{eq:erasure-inverse-trajectory}
  \widehat t'(p)=-\frac{1}{pR(\beta\lambda)},
  \qquad 
  \widehat t(1/2)=0,
  \qquad 
  \widehat t(\varepsilon)=\Gamma\tau.
\end{equation}
For $R'(\beta\lambda)\ne0$, this relation locally determines $\lambda$ as a function of $p$ and $\widehat t'$. 
The entropy production becomes
\begin{equation}\label{eq:erasure-inverse-functional}
  \Sigma[\widehat t]=\int_{\varepsilon}^{1/2}\mathcal L(p,\widehat t')\,\dd p,
  \qquad 
  \mathcal L(p,\widehat t')=\phi(\xi),
  \qquad 
  \phi(\xi):=\xi(1-\ee^{-\xi}).
\end{equation}
Since $\mathcal L$ has no explicit dependence on $\widehat t$, its Euler--Lagrange equation reduces to a first integral,
\begin{equation}\label{eq:erasure-first-integral}
  \frac{\dd}{\dd p}
  \frac{\partial\mathcal L}{\partial\widehat t'}=0,
  \qquad 
  \frac{\partial\mathcal L}{\partial\widehat t'}=\frac1C.
\end{equation}
At fixed $p$, the required derivatives are
\begin{equation}\label{eq:erasure-variational-derivatives}
  \frac{\partial\widehat t'}{\partial(\beta\lambda)}=\frac{R'(\beta\lambda)}{pR(\beta\lambda)^2},
  \qquad 
  \frac{\partial(\beta\lambda)}{\partial\widehat t'}=\frac{pR(\beta\lambda)^2}{R'(\beta\lambda)},
  \qquad 
  \phi'(\xi)=1-\ee^{-\xi}+\xi \ee^{-\xi}.
\end{equation}
The chain rule therefore yields
\begin{equation}\label{eq:erasure-optimal-equation}
  \frac1C=\frac{pR(\beta\lambda)^2}{R'(\beta\lambda)}(1-\ee^{-\xi}+\xi \ee^{-\xi}),
  \qquad
  CpR(\beta\lambda)(1-\ee^{-\xi}+\xi \ee^{-\xi})-\frac{R'(\beta\lambda)}{R(\beta\lambda)}=0.
\end{equation}
This is the optimality equation stated in the End Matter.

Solving Eq.~\eqref{eq:erasure-optimal-equation} gives the optimal energy gap $\lambda(p)$, with the constant $C$ fixed by
\begin{equation}\label{eq:erasure-time-matching}
  \Gamma\tau=\int_{\varepsilon}^{1/2}\frac{\dd p}{pR[\beta\lambda(p)]}.
\end{equation}
The time-dependent protocol is then obtained from Eq.~\eqref{eq:erasure-population}.
Substituting $\lambda(p)$ into Eq.~\eqref{eq:erasure-entropy-functional} yields the minimum entropy production,
\begin{equation}\label{eq:erasure-optimal-cost}
  \Sigma_{\min}(\tau,\varepsilon)=\int_{\varepsilon}^{1/2}\phi\!\left[\beta\lambda(p)-\ln\frac{1-p}{p}\right]\,\dd p.
\end{equation}
For finite $\tau^*$ and $\tau\geq\tau^*$, waiting for $\tau-\tau^*$ before following the protocol with $\xi=0$ gives $\Sigma_{\min}=0$. 
For $\tau<\tau^*$, the minimum entropy production is obtained by the procedure above.

\subsection{Collective information heat engine}
\label{sec:collective-engine}

\subsubsection{System--reservoir coupling and effective master equation}

We consider two ensembles $\alpha\in\{L,R\}$, each containing an odd number $N_s$ of two-level systems. 
The ensembles do not interact directly, and each exchanges particles with its own bosonic reservoir. 
The total Hamiltonian is
\begin{align}
  H_{\mathrm{tot}}&=H+\sum_{\alpha=L,R}H_{B,\alpha}+V,
  \qquad
  H=\sum_{\alpha=L,R}\left(\omega K_{\alpha,z}+\frac U2K_{\alpha,z}^2\right),
  \label{eq:collective-total-H}\\
  H_{B,\alpha}&=\sum_q\epsilon_{\alpha q}b_{\alpha q}^{\dagger}b_{\alpha q},
  \qquad
  V=\frac{1}{\sqrt{N_s}}\sum_{\alpha,q}\left(g_{\alpha q}K_{\alpha,+}b_{\alpha q}+g_{\alpha q}^{*}b_{\alpha q}^{\dagger}K_{\alpha,-}\right).
  \label{eq:collective-microscopic-coupling}
\end{align}
Here $K_{\alpha,z}=\frac12\sum_{n=1}^{N_s}\sigma_{\alpha,z}^{(n)}$, $K_{\alpha,\pm}=\sum_{n=1}^{N_s}\sigma_{\alpha,\pm}^{(n)}$, and $b_{\alpha q}$ annihilates a reservoir particle. 
The factor $N_s^{-1/2}$ normalizes the interaction amplitude and therefore contributes $N_s^{-1}$ to the transition rates. 
The system and reservoir number operators are
\begin{equation}\label{eq:collective-number-conservation}
  N_S=\sum_{\alpha=L,R}(K_{\alpha,z}+N_s/2),
  \qquad 
  N_{B,\alpha}=\sum_q b_{\alpha q}^{\dagger}b_{\alpha q},
  \qquad
  [H_{\mathrm{tot}},N_S+N_{B,L}+N_{B,R}]=0.
\end{equation}
We prepare each ensemble in the fully symmetric sector, whose Dicke states $|D_n^{N_s}\rangle_\alpha$ contain $n$ excitations, with $n=0,\ldots,N_s$.
The Dicke ladder and its energies satisfy
\begin{equation}\label{eq:collective-Dicke-ladder}
  \begin{aligned}
    K_{\alpha,+}|D_n^{N_s}\rangle_\alpha&=\sqrt{(N_s-n)(n+1)}\,|D_{n+1}^{N_s}\rangle_\alpha,\\
    E_n&=\omega\left(n-\frac{N_s}{2}\right)+\frac U2\left(n-\frac{N_s}{2}\right)^2.
  \end{aligned}
\end{equation}
The transition $n\to n+1$ requires energy $E_{n+1}-E_n=\omega+U[n-(N_s-1)/2]$. 
We assume that the reservoir couplings select the central transition at $\omega$ and suppress the other transitions, whose frequencies differ from $\omega$ by integer multiples of $|U|$. 
Starting in the central doublet then confines the dynamics to
\begin{equation}\label{eq:collective-central-doublet}
  |g\rangle_\alpha=|D_{(N_s-1)/2}^{N_s}\rangle_\alpha,
  \qquad
  |e\rangle_\alpha=|D_{(N_s+1)/2}^{N_s}\rangle_\alpha.
\end{equation}
The central matrix element and its normalized square are
\begin{equation}\label{eq:collective-enhancement}
  {}_\alpha\langle e|K_{\alpha,+}|g\rangle_\alpha=\frac{N_s+1}{2},
  \qquad
  \frac{K_{\alpha,+}}{\sqrt{N_s}}\longrightarrow\sqrt{\Lambda}\,|e\rangle_\alpha\langle g|,
  \qquad 
  \Lambda:=\frac{(N_s+1)^2}{4N_s}.
\end{equation}
Thus the collective enhancement is linear in $N_s$ for large ensembles with the normalization in Eq.~\eqref{eq:collective-microscopic-coupling}.

Both reservoirs have inverse temperature $\beta$. 
Under weak system--reservoir coupling, the usual Born--Markov approximation and secular approximation between distinct Bohr frequencies give a Lindblad equation for the two ensembles restricted to the states $|g\rangle_\alpha$ and $|e\rangle_\alpha$. 
Before feedback is applied, the reduced equation is
\begin{equation}\label{eq:collective-bare-master}
  \dot\rho_t=-i[H,\rho_t]+\sum_{\alpha=L,R}\sum_{k=0,1}\Diss{L_{\alpha,k}}[\rho_t],
  \qquad
  L_{\alpha,1}=\sqrt{\Lambda\gamma_{\alpha,1}}\,|e\rangle_\alpha\langle g|,
  \qquad
  L_{\alpha,0}=\sqrt{\Lambda\gamma_{\alpha,0}}\,|g\rangle_\alpha\langle e|.
\end{equation}
Each jump operator acts only on the corresponding ensemble.
The injection and ejection rates are
\begin{equation}\label{eq:collective-bosonic-rates}
  \gamma_{\alpha,1}=\Gamma_\alpha n_\alpha,
  \qquad 
  \gamma_{\alpha,0}=\Gamma_\alpha(n_\alpha+1),
  \qquad 
  n_\alpha=\frac{1}{\ee^{\beta(\omega-\mu_\alpha)}-1}.
\end{equation}
Here $\Gamma_\alpha>0$ sets the overall scale of the transition rates induced by reservoir $\alpha$, and $n_\alpha$ is its Bose occupation at energy $\omega$.

The chemical potentials enter through the reservoir states,
\begin{equation}\label{eq:collective-grand-canonical-reservoirs}
  \rho_{B,\alpha}=Z_\alpha^{-1}\ee^{-\beta(H_{B,\alpha}-\mu_\alpha N_{B,\alpha})},
  \qquad 
  \mu_L>\mu_R,
  \qquad 
  \mu_\alpha<\inf_q\epsilon_{\alpha q}.
\end{equation}
At fixed temperature, the chemical potential $\mu_\alpha$ is determined by the particle density of reservoir $\alpha$. 
The reservoir state above gives the Bose occupation in Eq.~\eqref{eq:collective-bosonic-rates}. 
The chemical potential lies below the lowest single-particle energy of the reservoir.
Each injection removes energy $\omega$ and one particle from reservoir $\alpha$, giving
\begin{equation}\label{eq:collective-local-detailed-balance}
  \frac{\gamma_{\alpha,1}}{\gamma_{\alpha,0}}=\ee^{-\beta(\omega-\mu_\alpha)},
  \qquad
  \Delta s_{\alpha,1}=-\beta(\omega-\mu_\alpha),
  \qquad 
  \Delta s_{\alpha,0}=\beta(\omega-\mu_\alpha).
\end{equation}

\subsubsection{Feedback dynamics and stationary state}

The two central doublets define the four states
\begin{equation}\label{eq:collective-four-states}
  |B\rangle=|g\rangle_L|g\rangle_R,
  \quad 
  |L\rangle=|e\rangle_L|g\rangle_R,
  \quad 
  |R\rangle=|g\rangle_L|e\rangle_R,
  \quad 
  |D\rangle=|e\rangle_L|e\rangle_R.
\end{equation}
After subtracting the fixed background energy and particle number,
\begin{equation}\label{eq:collective-effective-HN}
  H=\omega N,
  \qquad N=\proj{L}+\proj{R}+2\proj{D}.
\end{equation}
In particular, $|B\rangle$ contains no excitation relative to the two central ground states. 
It is not the empty state of all $2N_s$ sites. 
The bare jumps in this basis are
\begin{equation}\label{eq:collective-four-state-jumps}
  \begin{aligned}
    L_{L,1}&=\sqrt{\Lambda\gamma_{L,1}}(|L\rangle\langle B|+|D\rangle\langle R|),&L_{L,0}&=\sqrt{\Lambda\gamma_{L,0}}(|B\rangle\langle L|+|R\rangle\langle D|),\\
    L_{R,1}&=\sqrt{\Lambda\gamma_{R,1}}(|R\rangle\langle B|+|D\rangle\langle L|),&L_{R,0}&=\sqrt{\Lambda\gamma_{R,0}}(|B\rangle\langle R|+|L\rangle\langle D|).
  \end{aligned}
\end{equation}
The two terms in each jump operator describe transitions with the same energy change induced by the same reservoir.

After each detected injection, we apply
\begin{equation}\label{eq:collective-swap}
  U_{\mathrm{sw}}=\proj{B}+\proj{D}+|L\rangle\langle R|+|R\rangle\langle L|,
  \qquad
  [U_{\mathrm{sw}},H]=[U_{\mathrm{sw}},N]=0.
\end{equation}
No operation is applied after ejection. 
With ideal detection and negligible feedback delay, the resulting Wiseman equation is
\begin{equation}\label{eq:collective-feedback-master}
  \dot\rho_t=-i[H,\rho_t]+\sum_{\alpha=L,R}\left(\Diss{L_{\alpha,0}}[\rho_t]+\Diss{U_{\mathrm{sw}}L_{\alpha,1}}[\rho_t]\right).
\end{equation}
Equation~\eqref{eq:collective-swap} ensures that feedback exchanges neither energy nor particles with the system, so $P_F=0$.

A state diagonal in the four-state basis remains diagonal. 
Equation~\eqref{eq:collective-feedback-master} gives
\begin{equation}\label{eq:collective-populations}
  \begin{aligned}
    \Lambda^{-1}\dot p_B&=-(\gamma_{L,1}+\gamma_{R,1})p_B+\gamma_{L,0}p_L+\gamma_{R,0}p_R,\\
    \Lambda^{-1}\dot p_L&=\gamma_{R,1}p_B-(\gamma_{L,0}+\gamma_{R,1})p_L+\gamma_{R,0}p_D,\\
    \Lambda^{-1}\dot p_R&=\gamma_{L,1}p_B-(\gamma_{R,0}+\gamma_{L,1})p_R+\gamma_{L,0}p_D,\\
    \Lambda^{-1}\dot p_D&=\gamma_{R,1}p_L+\gamma_{L,1}p_R-(\gamma_{L,0}+\gamma_{R,0})p_D.
  \end{aligned}
\end{equation}
Define $\gamma_k:=\gamma_{L,k}+\gamma_{R,k}$. 
At stationarity, the last equation gives $\gamma_0p_D=\gamma_{R,1}p_L+\gamma_{L,1}p_R$.
Eliminating $p_D$ from the other equations yields
\begin{equation}\label{eq:collective-stationary-ratios}
  \frac{p_L}{p_B}=\frac{\gamma_{R,1}\gamma_0+\gamma_{L,1}\gamma_1}{\gamma_{L,0}(\gamma_0+\gamma_1)},
  \qquad
  \frac{p_R}{p_B}=\frac{\gamma_{L,1}\gamma_0+\gamma_{R,1}\gamma_1}{\gamma_{R,0}(\gamma_0+\gamma_1)}.
\end{equation}
Equivalently, the normalized solution is
\begin{equation}\label{eq:collective-stationary-state}
  \begin{aligned}
    p_B&=\frac{\gamma_{L,0}\gamma_{R,0}(\gamma_0+\gamma_1)}{\mathcal Z},&p_L&=\frac{\gamma_{R,0}(\gamma_{R,1}\gamma_0+\gamma_{L,1}\gamma_1)}{\mathcal Z},\\
    p_R&=\frac{\gamma_{L,0}(\gamma_{L,1}\gamma_0+\gamma_{R,1}\gamma_1)}{\mathcal Z},&p_D&=\frac{\gamma_{L,0}\gamma_{L,1}^2+\gamma_{R,0}\gamma_{R,1}^2+\gamma_{L,1}\gamma_{R,1}\gamma_1}{\mathcal Z},
  \end{aligned}
\end{equation}
where $\mathcal Z$ is the sum of the four numerators. 
The collective factor $\Lambda$ cancels from the stationary equations.

\subsubsection{Particle current, information consumption, and efficiency}

Let $J_N$ denote the net particle current entering from reservoir $R$. 
The bare jumps give
\begin{equation}\label{eq:collective-particle-current}
  J_N=\Lambda\left[\gamma_{R,1}(p_B+p_L)-\gamma_{R,0}(p_R+p_D)\right].
\end{equation}
We take particle transport from reservoir $R$ to reservoir $L$ as the positive direction and denote the currents through $|B\rangle$ and $|D\rangle$ by $J_B$ and $J_D$, respectively. 
The stationary conditions for $p_B$ and $p_D$ imply
\begin{equation}\label{eq:collective-branch-currents}
  \begin{aligned}
    J_B&:=\Lambda(\gamma_{R,1}p_B-\gamma_{R,0}p_R)=\Lambda(\gamma_{L,0}p_L-\gamma_{L,1}p_B),\\
    J_D&:=\Lambda(\gamma_{R,1}p_L-\gamma_{R,0}p_D)=\Lambda(\gamma_{L,0}p_D-\gamma_{L,1}p_R).
  \end{aligned}
\end{equation}
Using Eq.~\eqref{eq:collective-stationary-ratios}, we find
\begin{equation}\label{eq:collective-current-result}
  \begin{aligned}
    J_B&=\Lambda(\gamma_{R,1}-\gamma_{L,1})p_B\frac{\gamma_0}{\gamma_0+\gamma_1},
    \qquad
    J_D=\Lambda(\gamma_{R,1}-\gamma_{L,1})p_B\frac{\gamma_1}{\gamma_0+\gamma_1},\\
    J_N&=J_B+J_D=\Lambda(\gamma_{R,1}-\gamma_{L,1})p_B.
  \end{aligned}
\end{equation}
This proves the particle current expression in the main text. 
Both branches carry particles from $R$ to $L$ when $\gamma_{R,1}>\gamma_{L,1}$. 
Since $\mu_L>\mu_R$ implies $n_L>n_R$, engine operation requires $\Gamma_Rn_R>\Gamma_Ln_L$. 
This can be achieved by adjusting the reservoir coupling strengths.

Each particle transported from $R$ to $L$ delivers chemical work $\Delta\mu:=\mu_L-\mu_R>0$. 
The stationary heat currents released into the reservoirs are
\begin{equation}\label{eq:collective-reservoir-currents}
  \dot Q_R=-(\omega-\mu_R)J_N,
  \qquad
  \dot Q_L=(\omega-\mu_L)J_N.
\end{equation}
Thus
\begin{equation}\label{eq:collective-output-entropy-flow}
  P=\Delta\mu\,J_N=-\dot Q_L-\dot Q_R,
  \qquad 
  \dot S_{\mathrm{env}}=-\beta P.
\end{equation}
Without feedback, each ensemble relaxes independently to its own equilibrium state, and both stationary reservoir currents vanish.

We next evaluate the information rate. 
An injection from $R$ occurs at rate $a_{R,1}=\Lambda\gamma_{R,1}(p_B+p_L)$ and prepares
\begin{equation}\label{eq:collective-conditional-state}
  \rho'_{R,1}=\frac{p_B\proj{R}+p_L\proj{D}}{p_B+p_L},
  \qquad
  U_{\mathrm{sw}}\rho'_{R,1}U_{\mathrm{sw}}^{\dagger}=\frac{p_B\proj{L}+p_L\proj{D}}{p_B+p_L}.
\end{equation}
The unitary preserves the conditional-state entropy. 
Its contribution to Eq.~\eqref{eq:information-rate} is therefore
\begin{equation}\label{eq:collective-information-contribution}
  a_{R,1}\left[D(U_{\mathrm{sw}}\rho'_{R,1}U_{\mathrm{sw}}^{\dagger}\Vert\rho)-D(\rho'_{R,1}\Vert\rho)\right]=\Lambda\gamma_{R,1}p_B\ln\frac{p_R}{p_L}.
\end{equation}
Exchanging $L$ and $R$ gives the contribution of injection from $L$. 
Ejections have no feedback, and the component in $|D\rangle$ is invariant under the swap. 
Summing gives
\begin{equation}\label{eq:collective-information-rate}
  \dot I=\Lambda(\gamma_{R,1}-\gamma_{L,1})p_B\ln\frac{p_R}{p_L}=J_N\ln\frac{p_R}{p_L}.
\end{equation}
Local detailed balance gives $\beta\Delta\mu=\ln[\gamma_{L,1}\gamma_{R,0}/(\gamma_{R,1}\gamma_{L,0})]$. 
Thus
\begin{equation}\label{eq:collective-affinity}
  \begin{split}
    \ln\frac{p_L}{p_R}-\beta\Delta\mu&=\ln\frac{\gamma_{R,1}\gamma_{L,0}p_L}{\gamma_{L,1}\gamma_{R,0}p_R}\\
    &=\ln\left[1+\frac{(\gamma_{R,1}-\gamma_{L,1})\gamma_0\gamma_1}{\gamma_{L,1}(\gamma_{L,1}\gamma_0+\gamma_{R,1}\gamma_1)}\right],
  \end{split}
\end{equation}
where the last equality follows from Eq.~\eqref{eq:collective-stationary-ratios}.
Hence $\ln(p_L/p_R)>\beta\Delta\mu>0$ for $\gamma_{R,1}>\gamma_{L,1}$.
Equation~\eqref{eq:collective-information-rate} then implies $\dot I<0$. 
Since feedback is unitary and the system entropy is constant, we have
\begin{equation}\label{eq:collective-entropy-efficiency}
  \sigma_{\mathrm{ss}}=-\beta P-\dot I=J_N\left[\ln\frac{p_L}{p_R}-\beta\Delta\mu\right],
  \qquad
  \eta_I=\frac{P}{-\dot I/\beta}=\frac{\beta\Delta\mu}{\ln(p_L/p_R)}.
\end{equation}
These relations establish the information consumption and efficiency used in the main text.

To evaluate the bound, we choose the following weights representing the same stationary output power:
\begin{equation}\label{eq:collective-work-weights}
  c_{\alpha k}^{xy}=\frac{\Delta\mu\,f_{\alpha k}^{xy}}{\ln(p_L/p_R)-\beta\Delta\mu}.
\end{equation}
Since $\frac12\sum_{\alpha,k,x,y}j_{\alpha k}^{xy}f_{\alpha k}^{xy}=J_N[\ln(p_L/p_R)-\beta\Delta\mu]$, these weights give $J_{\mathrm{ss}}=\Delta\mu J_N=P$. 
Substituting them into Eq.~\eqref{eq:kinetic-coefficient} yields the weighted mobility
\begin{equation}\label{eq:collective-optimal-coefficient}
  m_{\mathrm{ss}}=\frac{(\Delta\mu)^2J_N}{\ln(p_L/p_R)-\beta\Delta\mu}.
\end{equation}
Together with Eq.~\eqref{eq:collective-entropy-efficiency}, this gives
\begin{equation}\label{eq:collective-bound-saturation}
  P=\beta m_{\mathrm{ss}}\frac{1-\eta_I}{\eta_I},
\end{equation}
showing that the power--efficiency bound can be saturated.

\subsubsection{Finite power as the information efficiency approaches unity}

As in the main text, we keep $\beta$, $\omega$, $\mu_L$, $\mu_R$, and $\gamma_{L,1}>0$ fixed and choose, in the same rate units,
\begin{equation}\label{eq:collective-reversible-family}
  \gamma_{R,1}=\gamma_{L,1}+\frac1{N_s}.
\end{equation}
As $N_s\to\infty$, the injection rates become equal. 
Setting $\gamma_{R,1}=\gamma_{L,1}$ in Eq.~\eqref{eq:collective-stationary-state} and using local detailed balance gives
\begin{equation*}
  \begin{gathered}
    \frac{p_L}{p_B}=\ee^{-\beta(\omega-\mu_L)},
    \qquad
    \frac{p_R}{p_B}=\ee^{-\beta(\omega-\mu_R)},\\
    \frac{p_D}{p_B}=\ee^{-\beta(\omega-\mu_L)}\ee^{-\beta(\omega-\mu_R)}.
  \end{gathered}
\end{equation*}
Normalization, $p_B+p_L+p_R+p_D=1$, determines the limiting population $p_B^{\mathrm{rev}}$. 
Since the stationary solution depends smoothly on the rates, expanding about $\gamma_{R,1}=\gamma_{L,1}$ gives
\begin{equation}\label{eq:collective-reversible-state}
  p_B=p_B^{\mathrm{rev}}+O(N_s^{-1}),
  \qquad
  p_B^{\mathrm{rev}}=\prod_{\alpha=L,R}[1+\ee^{-\beta(\omega-\mu_\alpha)}]^{-1}.
\end{equation}
Using $\Lambda=N_s/4+O(1)$ in Eq.~\eqref{eq:collective-current-result}, we obtain
\begin{equation}\label{eq:collective-finite-power-limit}
  J_N=\frac{p_B^{\mathrm{rev}}}{4}+O(N_s^{-1}),
  \qquad
  P\longrightarrow\frac{\Delta\mu}{4}p_B^{\mathrm{rev}}>0.
\end{equation}
Substituting Eq.~\eqref{eq:collective-reversible-family} into Eq.~\eqref{eq:collective-affinity} gives
\begin{equation}\label{eq:collective-affinity-expansion}
  \ln\frac{p_L}{p_R}-\beta\Delta\mu=\ln\left[1+\frac1{N_s}\frac{\gamma_0\gamma_1}{\gamma_{L,1}(\gamma_{L,1}\gamma_0+\gamma_{R,1}\gamma_1)}\right].
\end{equation}
The fraction multiplying $1/N_s$ tends to a finite positive value. 
Using $\ln(1+x)=x+O(x^2)$, the left-hand side therefore scales as $1/N_s$. 
Since $\ln(p_L/p_R)\to\beta\Delta\mu>0$ and $J_N\to p_B^{\mathrm{rev}}/4>0$, Eqs.~\eqref{eq:collective-entropy-efficiency} and \eqref{eq:collective-optimal-coefficient} give
\begin{equation}\label{eq:collective-reversible-scalings}
  1-\eta_I=O(N_s^{-1}),
  \qquad 
  m_{\mathrm{ss}}\propto N_s,
  \qquad 
  \sigma_{\mathrm{ss}}=O(N_s^{-1}).
\end{equation}
The collective enhancement keeps the output power finite as $\sigma_{\mathrm{ss}}\to0$.

\subsection{A three-state information heat pump}
\label{sec:three-state-heat-pump}

We illustrate the heating-power--COP bound in Sec.~\ref{sec:heat-pump-bound} with a three-state system in which feedback transfers population between two degenerate excited states. 
The model allows the stationary heat currents, information consumption, and weighted mobilities to be evaluated explicitly.

\subsubsection{Model and stationary state}

The system Hamiltonian is
\begin{equation}\label{eq:pump-example-H}
  H=\omega(\proj{c}+\proj{h}),\qquad \omega>0,
\end{equation}
where $|0\rangle$ is the ground state. 
The cold reservoir, at inverse temperature $\beta_c$, couples only to $|0\rangle\leftrightarrow|c\rangle$, and the hot reservoir, at $\beta_h<\beta_c$, couples only to $|0\rangle\leftrightarrow|h\rangle$. 
A corresponding exchange interaction with independent bosonic reservoirs is
\begin{equation}\label{eq:pump-example-coupling}
  V=\sum_{\nu=c,h}\sum_q\left(g_{\nu q}|\nu\rangle\langle0|b_{\nu q}+g_{\nu q}^{*}b_{\nu q}^{\dagger}|0\rangle\langle\nu|\right).
\end{equation}
Under the Born--Markov and secular approximations, the bare absorption and emission jumps have the form
\begin{equation}\label{eq:pump-example-jumps}
  L_{\nu,1}=\sqrt{\gamma_{\nu,1}}\,|\nu\rangle\langle0|,
  \qquad
  L_{\nu,0}=\sqrt{\gamma_{\nu,0}}\,|0\rangle\langle\nu|,
  \qquad
  \frac{\gamma_{\nu,1}}{\gamma_{\nu,0}}=\ee^{-\beta_\nu\omega},
  \quad \nu\in\{c,h\}.
\end{equation}
All four rate coefficients are positive. After either absorption jump, the controller applies the unitary
\begin{equation}\label{eq:pump-example-swap}
  U_{\mathrm{sw}}=\proj{0}+|c\rangle\langle h|+|h\rangle\langle c|,
  \qquad 
  [U_{\mathrm{sw}},H]=0.
\end{equation}
No feedback is applied after emission. 
With ideal detection and negligible delay, the master equation is
\begin{equation}\label{eq:pump-example-master}
  \dot\rho_t=-i[H,\rho_t]+\sum_{\nu=c,h}\left(\Diss{U_{\mathrm{sw}}L_{\nu,1}}[\rho_t]+\Diss{L_{\nu,0}}[\rho_t]\right).
\end{equation}
In particular, $U_{\mathrm{sw}}L_{c,1}=\sqrt{\gamma_{c,1}}|h\rangle\langle0|$ and $U_{\mathrm{sw}}L_{h,1}=\sqrt{\gamma_{h,1}}|c\rangle\langle0|$. 
Thus absorption from the cold reservoir prepares $|h\rangle$ after feedback, from which emission returns energy to the hot reservoir. 
The reverse heat transfer sequence is also allowed. Since the swap commutes with $H$, its energy input vanishes for every state: $P_F=\Tr(H\cB[\rho])=0$.

For $\rho=p_0\proj{0}+p_c\proj{c}+p_h\proj{h}$, the population equations are
\begin{equation}\label{eq:pump-example-populations}
  \begin{aligned}
    \dot p_0&=-(\gamma_{c,1}+\gamma_{h,1})p_0+\gamma_{c,0}p_c+\gamma_{h,0}p_h,\\
    \dot p_c&=\gamma_{h,1}p_0-\gamma_{c,0}p_c,\\
    \dot p_h&=\gamma_{c,1}p_0-\gamma_{h,0}p_h.
  \end{aligned}
\end{equation}
Setting the last two equations to zero and imposing normalization gives
\begin{equation}\label{eq:pump-example-stationary-state}
  p_c=\frac{\gamma_{h,1}}{\gamma_{c,0}}p_0,
  \qquad
  p_h=\frac{\gamma_{c,1}}{\gamma_{h,0}}p_0,
  \qquad
  p_0=\left(1+\frac{\gamma_{h,1}}{\gamma_{c,0}}
  +\frac{\gamma_{c,1}}{\gamma_{h,0}}\right)^{-1}.
\end{equation}
All coherences decay under Eq.~\eqref{eq:pump-example-master}.

\subsubsection{Heat currents and information consumption}

Evaluating the reservoir heat currents in the stationary state of Eq.~\eqref{eq:pump-example-stationary-state} gives
\begin{equation}\label{eq:pump-example-heat-currents}
  \begin{aligned}
    \dot Q_c&=\omega(\gamma_{c,0}p_c-\gamma_{c,1}p_0)=-\omega(\gamma_{c,1}-\gamma_{h,1})p_0,\\
    \dot Q_h&=\omega(\gamma_{h,0}p_h-\gamma_{h,1}p_0)=\omega(\gamma_{c,1}-\gamma_{h,1})p_0.
  \end{aligned}
\end{equation}
Define the stationary heat transfer rate
\begin{equation}\label{eq:pump-example-transfer-rate}
  J:=(\gamma_{c,1}-\gamma_{h,1})p_0.
\end{equation}
The heat pump regime is $\gamma_{c,1}>\gamma_{h,1}$, for which $\dot Q_c=-\omega J<0$ and $\dot Q_h=\omega J>0$. 
The power is consequently
\begin{equation}\label{eq:pump-example-heating-power}
  P_{\mathrm{hp}}=\dot Q_h=\omega J=\frac{\omega(\gamma_{c,1}-\gamma_{h,1})}{1+\gamma_{h,1}/\gamma_{c,0}+\gamma_{c,1}/\gamma_{h,0}}.
\end{equation}
The extracted cold heat equals the heat delivered to the hot reservoir, consistent with $P_F=0$.

A cold bath absorption prepares $\proj{c}$ before feedback and $\proj{h}$ after feedback. 
Since $D(\proj{i}\Vert\rho)=-\ln p_i$, its information contribution is
\begin{equation}\label{eq:pump-example-cold-information}
  \dot I_c=\gamma_{c,1}p_0\left[D(\proj{h}\Vert\rho)-D(\proj{c}\Vert\rho)\right]=\gamma_{c,1}p_0\ln\frac{p_c}{p_h}.
\end{equation}
The hot-bath contribution follows by exchanging $c$ and $h$. 
Emission events do not change the information through feedback. Hence
\begin{equation}\label{eq:pump-example-information-rate}
  \dot I=\dot I_c+\dot I_h=(\gamma_{c,1}-\gamma_{h,1})p_0\ln\frac{p_c}{p_h}=J\ln\frac{p_c}{p_h}.
\end{equation}
Detailed balance and Eq.~\eqref{eq:pump-example-stationary-state} give
\begin{equation}\label{eq:pump-example-population-ratio}
  \ln\frac{p_h}{p_c}=(\beta_c-\beta_h)\omega+2\ln\frac{\gamma_{c,1}}{\gamma_{h,1}}>0.
\end{equation}
Thus $\dot I=-J\ln(p_h/p_c)<0$. 
Since feedback is unitary and the system entropy is constant, the entropy-production rate and COP are
\begin{equation}\label{eq:pump-example-entropy-cop}
  \sigma_{\mathrm{ss}}=\sigma_{\mathrm{ss}}^{(0)}=2J\ln\frac{\gamma_{c,1}}{\gamma_{h,1}},
  \qquad
  \kappa=\frac{\beta_c\omega}{\ln(p_h/p_c)}<\kappa_C.
\end{equation}

The forward and reverse stationary fluxes are $\gamma_{c,1}p_0$ and $\gamma_{h,1}p_0$ for the cold reservoir, and occur in the opposite order for the hot reservoir.
Their logarithmic means are therefore identical. 
Equations~\eqref{eq:heat-pump-bath-coefficients} and \eqref{eq:heat-pump-effective-coefficient} give
\begin{equation}\label{eq:pump-example-heat-coefficients}
  m_{Q_c}=m_{Q_h}=\frac{\omega^2J}{\ln(\gamma_{c,1}/\gamma_{h,1})},
  \qquad
  m_{\mathrm{hp}}=\frac{\omega^2J}{2\ln(\gamma_{c,1}/\gamma_{h,1})}.
\end{equation}
Together with Eq.~\eqref{eq:pump-example-entropy-cop}, this yields
\begin{equation}\label{eq:pump-example-bound-saturation}
  \beta_c m_{\mathrm{hp}}\left(\frac1\kappa-\frac1{\kappa_C}\right)=\omega J=P_{\mathrm{hp}},
\end{equation}
showing that Eq.~\eqref{eq:heat-pump-energy-neutral-bound} is saturated throughout the heat-pump regime.

\subsubsection{Numerical illustration}

\begin{figure}[!htbp]
  \centering
  \includegraphics[width=0.68\linewidth]{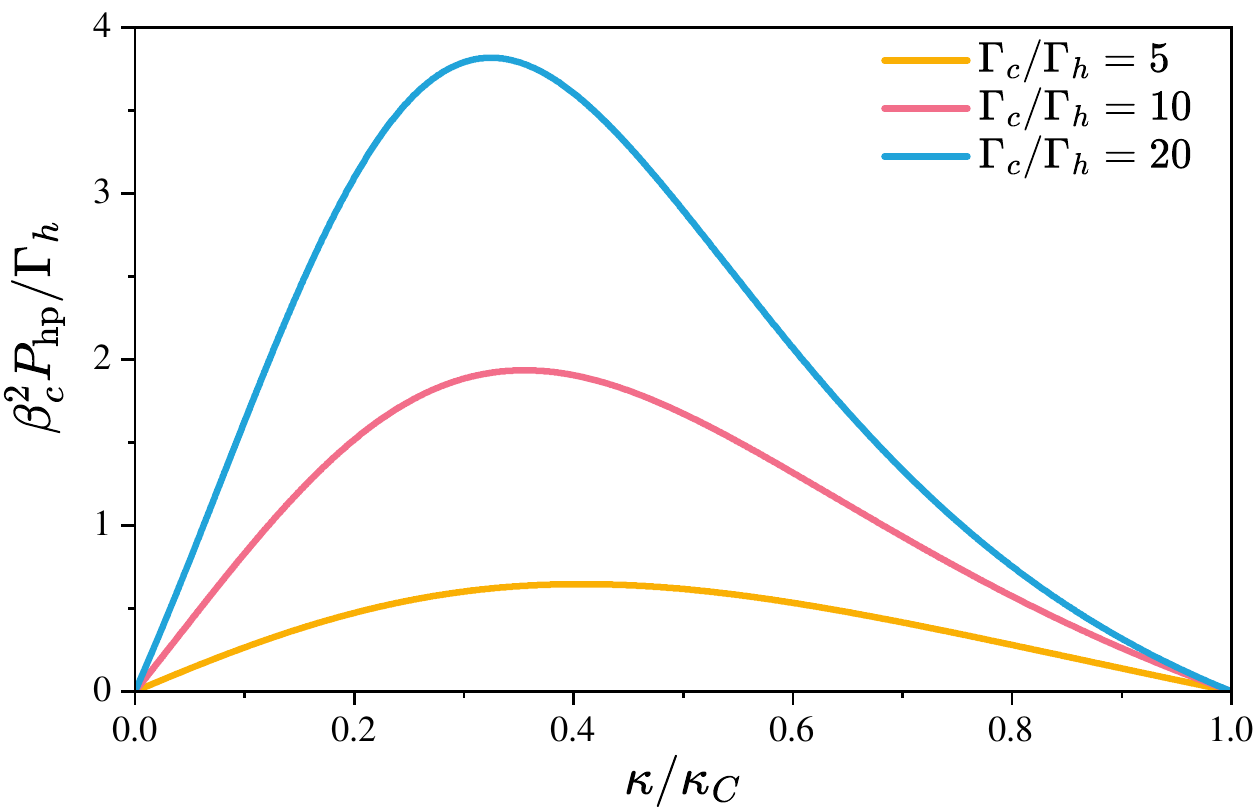}
  \caption{Heating power versus normalized COP for the three-state information heat pump. The orange, pink, and blue curves correspond to $\Gamma_c/\Gamma_h=5,10,20$, respectively. We use the Ohmic spectral density in Eq.~\eqref{eq:pump-example-ohmic-rates}, with $\beta_c=1$, $\beta_h=1/2$, and $\Gamma_h=1$, and vary the excitation gap over $0<\omega<2\ln(\Gamma_c/\Gamma_h-1)$.}
  \label{fig:three-state-heat-pump}
\end{figure}

For Fig.~\ref{fig:three-state-heat-pump}, we use bosonic reservoirs with Ohmic spectral densities, giving the rates
\begin{equation}\label{eq:pump-example-ohmic-rates}
  \gamma_{\nu,1}=\frac{\Gamma_\nu\omega}{\ee^{\beta_\nu\omega}-1},
  \qquad
  \gamma_{\nu,0}=\frac{\Gamma_\nu\omega}{1-\ee^{-\beta_\nu\omega}},
  \qquad \nu\in\{c,h\}.
\end{equation}
We set $\beta_c=1$, $\beta_h=1/2$, and $\Gamma_h=1$, and take $R_\Gamma:=\Gamma_c/\Gamma_h=5,10,20$. 
For each coupling ratio, we scan $0<\omega<2\ln(R_\Gamma-1)$ and evaluate the heating power and COP from Eqs.~\eqref{eq:pump-example-heating-power} and \eqref{eq:pump-example-entropy-cop}. 
The figure plots $\beta_c^2P_{\mathrm{hp}}/\Gamma_h$ against $\kappa/\kappa_C$. 
Each curve saturates the bound in Eq.~\eqref{eq:heat-pump-energy-neutral-bound}.

\FloatBarrier

\subsection{A finite-time feedback Otto engine}
\label{sec:feedback-otto-example}

\begin{figure}[!t]
  \centering
  \includegraphics[width=0.68\linewidth]{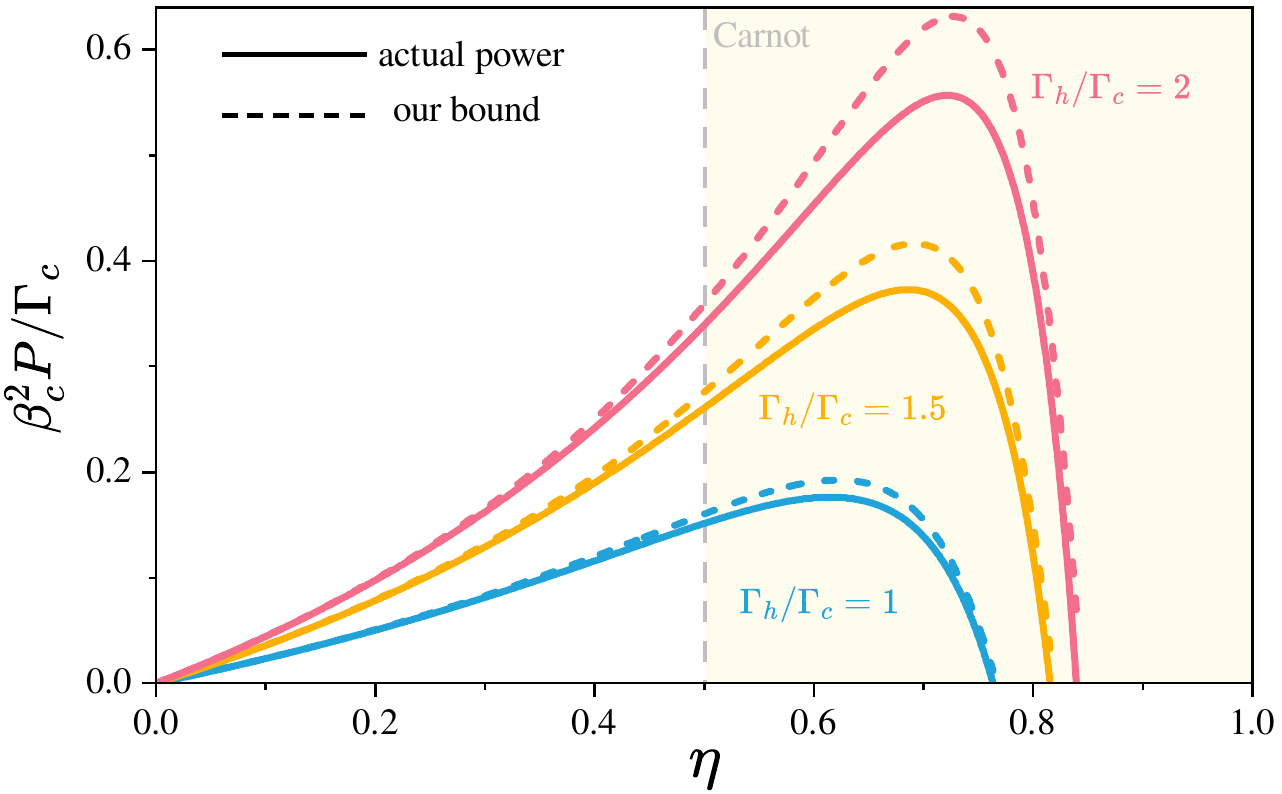}
  \caption{Output power versus thermal efficiency $\eta=W/Q_h$ for the feedback Otto engine. Solid curves show the power, and dashed curves show the bound Eq.~\eqref{eq:otto-quadratic-bound}, evaluated along the same periodic trajectories. The blue, orange, and pink curves correspond to $\Gamma_h/\Gamma_c=1,1.5,2$, respectively. We use the Ohmic spectral density in Eq.~\eqref{eq:otto-ohmic-rates}, with $\beta_c=1$, $\beta_h=1/2$, $\omega_c=\Gamma_c=1$, and $t_h=t_c=0.1$, and vary $\omega_h=\omega_c/(1-\eta)$. The vertical line marks $\eta_C=1/2$. In the shaded region, information consumption allows positive power at $\eta>\eta_C$.}
  \label{fig:feedback-otto}
\end{figure}

We illustrate the power--efficiency bound for cyclic heat engines derived in Sec.~\ref{sec:two-reservoir-engine} using an Otto cycle with feedback. 
The working medium has a ground state $|0\rangle$ and two degenerate excited states $|h\rangle$ and $|c\rangle$. 
Feedback transfers an absorbed excitation to the state that is decoupled from the active reservoir. 
Work is extracted by lowering the excitation gap during the expansion stroke. 
The power--efficiency curves are shown in Fig.~\ref{fig:feedback-otto}.

\subsubsection{Cycle and feedback dynamics}

Let
\begin{equation}\label{eq:otto-H}
  \Pi_e:=\proj{h}+\proj{c},
  \qquad
  H_\nu=\omega_\nu\Pi_e,
  \qquad
  \omega_h>\omega_c>0.
\end{equation}
During contact with reservoir $\nu\in\{h,c\}$, the Hamiltonian is fixed at $H_\nu$, and the other reservoir is disconnected. 
Reservoir $\nu$ couples only the transition $|0\rangle\leftrightarrow|\nu\rangle$, with bare absorption and emission jumps
\begin{equation}\label{eq:otto-jumps}
  L_{\nu,1}=\sqrt{\gamma_{\nu,1}}\,|\nu\rangle\langle0|,
  \qquad
  L_{\nu,0}=\sqrt{\gamma_{\nu,0}}\,|0\rangle\langle\nu|,
  \qquad
  \frac{\gamma_{\nu,1}}{\gamma_{\nu,0}}=\ee^{-\beta_\nu\omega_\nu}.
\end{equation}
These jumps follow from an exchange coupling of the form in Eq.~\eqref{eq:pump-example-coupling}, with only reservoir $\nu$ coupled and its transition frequency equal to $\omega_\nu$. 
After either absorption, the controller applies
\begin{equation}\label{eq:otto-swap}
  U_{\mathrm{sw}}=\proj{0}+|h\rangle\langle c|+|c\rangle\langle h|,
  \qquad 
  [U_{\mathrm{sw}},H_h]=[U_{\mathrm{sw}},H_c]=0.
\end{equation}
No feedback is applied after emission. 
The isochoric generators are
\begin{equation}\label{eq:otto-generators}
  \cL_\nu^{(\mathrm{fb})}[\rho]=-i[H_\nu,\rho]+\Diss{U_{\mathrm{sw}}L_{\nu,1}}[\rho]+\Diss{L_{\nu,0}}[\rho].
\end{equation}
Every feedback event conserves energy, giving $P_F=0$ and $E_F=0$.

After hot absorption, feedback transfers the excitation to $|c\rangle$, which is uncoupled from the hot reservoir, so that the excitation can contribute work during the subsequent expansion.

The cycle consists of four strokes:
\begin{enumerate}
  \item \textit{Compression.} With both reservoirs disconnected, the excitation gap is increased from $\omega_c$ to $\omega_h$, changing the Hamiltonian from $H_c$ to $H_h$.
  \item \textit{Hot isochore.} The system is coupled only to the hot reservoir and evolves under $\cL_h^{(\mathrm{fb})}$ for a time $t_h$, with the Hamiltonian fixed at $H_h$. Each absorption is followed by the swap $U_{\mathrm{sw}}$.
  \item \textit{Expansion.} With both reservoirs disconnected, the excitation gap is decreased from $\omega_h$ to $\omega_c$, changing the Hamiltonian from $H_h$ to $H_c$ and extracting work.
  \item \textit{Cold isochore.} The system is coupled only to the cold reservoir and evolves under $\cL_c^{(\mathrm{fb})}$ for a time $t_c$, with the Hamiltonian fixed at $H_c$. Each absorption is followed by the swap $U_{\mathrm{sw}}$.
\end{enumerate}
All Hamiltonians have the same eigenbasis, and the state remains diagonal in this basis. 
The gap changes therefore preserve the populations without generating coherence. 
We neglect the durations of the compression and expansion strokes, so the cycle time is $\tau=t_h+t_c$.

Writing $\rho=p_0\proj{0}+p_h\proj{h}+p_c\proj{c}$, the population equations are
\begin{equation}\label{eq:otto-populations}
  \begin{aligned}
    \text{hot:}\qquad\dot p_0&=-\gamma_{h,1}p_0+\gamma_{h,0}p_h,
    &\dot p_h&=-\gamma_{h,0}p_h,
    &\dot p_c&=\gamma_{h,1}p_0,\\
    \text{cold:}\qquad\dot p_0&=-\gamma_{c,1}p_0+\gamma_{c,0}p_c,
    &\dot p_h&=\gamma_{c,1}p_0,
    &\dot p_c&=-\gamma_{c,0}p_c.
  \end{aligned}
\end{equation}
We set $t=0$ at the beginning of the hot isochore. 
The hot and cold contacts occupy $0\leq t\leq t_h$ and $t_h\leq t\leq\tau$, respectively. 
Solving Eq.~\eqref{eq:otto-populations} gives
\begin{equation}\label{eq:otto-bright-population}
  \begin{aligned}
    p_h(t)&=\ee^{-\gamma_{h,0}t}p_h(0),&&0\leq t\leq t_h,\\
    p_c(t)&=\ee^{-\gamma_{c,0}(t-t_h)}p_c(t_h),&&t_h\leq t\leq\tau,
  \end{aligned}
\end{equation}
\begin{equation}\label{eq:otto-ground-propagator}
  p_0(t)=
  \begin{cases}
    \ee^{-\gamma_{h,1}t}p_0(0)+\dfrac{\gamma_{h,0}\left(\ee^{-\gamma_{h,1}t}-\ee^{-\gamma_{h,0}t}\right)}{\gamma_{h,0}-\gamma_{h,1}}p_h(0),&0\leq t\leq t_h,\\[8pt]
    \ee^{-\gamma_{c,1}(t-t_h)}p_0(t_h)+\dfrac{\gamma_{c,0}\left(\ee^{-\gamma_{c,1}(t-t_h)}-\ee^{-\gamma_{c,0}(t-t_h)}\right)}{\gamma_{c,0}-\gamma_{c,1}}p_c(t_h),&t_h\leq t\leq\tau.
  \end{cases}
\end{equation}
The remaining population follows from normalization. 
Applying these solutions to the hot and cold isochores and imposing $p_i(\tau)=p_i(0)$ for $i\in\{0,h,c\}$ determines the periodic state. 
Positive rates and contact times give a unique positive periodic state.

Without the swaps, the cold stroke leaves $p_h$ unchanged and the hot stroke leaves $p_c$ unchanged. 
Periodicity then forces the net excitation change on each contact to vanish, giving zero cycle heat and work for this model.

\subsubsection{Heat, work, and information per cycle}

Define the total population of excited states as $n(t):=p_h(t)+p_c(t)$. 
Its change during the hot isochore is $\Delta n:=n(t_h)-n(0)$. 
Since the Hamiltonian is fixed and feedback conserves energy, the heat absorbed during this stroke is $\omega_h\Delta n$. 
Periodicity requires the population change during the cold isochore to be $-\Delta n$. 
The absorbed hot heat and released cold heat are therefore
\begin{equation}\label{eq:otto-cycle-heats}
  Q_h=-\int_0^{t_h}\dot Q_h(t)\,\dd t=\omega_h\Delta n,
  \qquad
  Q_c=\int_{t_h}^{\tau}\dot Q_c(t)\,\dd t=\omega_c\Delta n.
\end{equation}
Here $Q_c$ is the positive heat released to the cold reservoir, following Sec.~\ref{sec:two-reservoir-engine}. 
The expansion delivers $(\omega_h-\omega_c)n(t_h)$, while compression requires $(\omega_h-\omega_c)n(0)$. 
Hence
\begin{equation}\label{eq:otto-work-efficiency}
  W=(\omega_h-\omega_c)\Delta n=Q_h-Q_c,
  \qquad
  P=\frac{W}{\tau},
  \qquad
  \eta=\frac{W}{Q_h}=1-\frac{\omega_c}{\omega_h}.
\end{equation}
The engine regime has $\Delta n>0$. 
The heat engine efficiency $\eta$ uses only the absorbed hot heat as its input; the consumed information enters the entropy balance separately. 
Using an efficiency that does not include information consumption facilitates comparison with power--efficiency bounds for heat engines without feedback. 
An efficiency that includes the information contribution can also be defined, but we do not discuss it here.

A hot absorption prepares $\proj{h}$ and is followed by a swap to $\proj{c}$. 
Since $D(\proj{i}\Vert\rho)=-\ln p_i$, its information contribution is $\gamma_{h,1}p_0\ln(p_h/p_c)$. 
The cold contribution follows by exchanging $h$ and $c$. 
Thus
\begin{equation}\label{eq:otto-cycle-information}
  \dot I_h=\gamma_{h,1}p_0\ln\frac{p_h}{p_c},
  \qquad
  \dot I_c=\gamma_{c,1}p_0\ln\frac{p_c}{p_h},
  \qquad
  I=\int_0^{t_h}\dot I_h\,\dd t+\int_{t_h}^{\tau}\dot I_c\,\dd t.
\end{equation}
There is no information contribution on the work strokes. 
For the parameter families shown in Fig.~\ref{fig:feedback-otto}, $I<0$. 
Define
\begin{equation}\label{eq:otto-information-power}
  P_I:=-\frac{I}{\beta_c\tau}>0.
\end{equation}
Because the feedback is unitary, $\Sigma_\tau=\Sigma_\tau^{(0)}$. 
The cycle entropy production is
\begin{equation}\label{eq:otto-cycle-entropy}
  \Sigma_\tau=\beta_cQ_c-\beta_hQ_h-I=\beta_c\tau\left[\frac{P}{\eta}(\eta_C-\eta)+P_I\right],
  \qquad
  \eta_C=1-\frac{\beta_h}{\beta_c}.
\end{equation}
For $\eta>\eta_C$, nonnegative entropy production requires $P_I\geq P(\eta-\eta_C)/\eta$. 
The power bound derived below further constrains the required information consumption.

\subsubsection{Power--efficiency bound}

For the combined current $\dot Q_c-\dot Q_h$, we assign weights $+\omega_h$ and $-\omega_h$ to hot absorption and emission, and $-\omega_c$ and $+\omega_c$ to cold absorption and emission, respectively. 
Since the weights enter the weighted mobility quadratically, each reservoir contributes the squared gap $\omega_\nu^2$. 
Substituting the absorption and emission fluxes $\gamma_{\nu,1}p_0(t)$ and $\gamma_{\nu,0}p_\nu(t)$ into Eq.~\eqref{eq:two-reservoir-average-coefficients} gives
\begin{equation}\label{eq:otto-instantaneous-coefficients}
  m_{Q_\nu}(t)=\omega_\nu^2\frac{\gamma_{\nu,1}p_0(t)-\gamma_{\nu,0}p_\nu(t)}{\ln\!\left[\gamma_{\nu,1}p_0(t)/\bigl(\gamma_{\nu,0}p_\nu(t)\bigr)\right]}.
\end{equation}
The ratio is the logarithmic mean of the two fluxes, defined by continuity when they coincide. 
Only one reservoir is coupled at a time, so the cycle-averaged mobility entering the power bound is
\begin{equation}\label{eq:otto-cycle-coefficients}
  \overline{m}=\frac1\tau\left[\int_0^{t_h}m_{Q_h}(t)\,\dd t+\int_{t_h}^{\tau}m_{Q_c}(t)\,\dd t\right].
\end{equation}
Since $E_F=0$, the resource ratios in Eq.~\eqref{eq:two-reservoir-resource-ratios} satisfy $\zeta_F=0$ and $\zeta=\zeta_I=\eta P_I/P$. 
Substituting into Eq.~\eqref{eq:two-reservoir-combined-power-bound} and multiplying by $P>0$ yields
\begin{equation}\label{eq:otto-quadratic-bound}
  P^2\leq\frac{\beta_c\overline{m}}{(2-\eta)^2}\left[\eta(\eta_C-\eta)P+\eta^2P_I\right].
\end{equation}
The power bound shown in Fig.~\ref{fig:feedback-otto} is the positive root of the corresponding quadratic equation. 
Here $P_I$ and $\overline{m}$ are evaluated along the periodic trajectory at each operating point.

\subsubsection{Numerical illustration}

For Fig.~\ref{fig:feedback-otto}, we use bosonic reservoirs with Ohmic rates,
\begin{equation}\label{eq:otto-ohmic-rates}
  \gamma_{\nu,1}=\frac{\Gamma_\nu\omega_\nu}{\ee^{\beta_\nu\omega_\nu}-1},
  \qquad
  \gamma_{\nu,0}=\frac{\Gamma_\nu\omega_\nu}{1-\ee^{-\beta_\nu\omega_\nu}},\qquad \nu\in\{h,c\}.
\end{equation}
We set $\beta_c=1$, $\beta_h=1/2$, $\omega_c=\Gamma_c=1$, and $t_h=t_c=0.1$, and take $r_\Gamma:=\Gamma_h/\Gamma_c=1,1.5,2$. 
For each coupling ratio, we vary $\omega_h=\omega_c/(1-\eta)$ along the engine regime. 
We determine the periodic state using Eqs.~\eqref{eq:otto-bright-population} and \eqref{eq:otto-ground-propagator}, evaluate the power from
Eq.~\eqref{eq:otto-work-efficiency}, and calculate $P_I$ and $\overline{m}$ from Eqs.~\eqref{eq:otto-cycle-information}, \eqref{eq:otto-information-power}, and \eqref{eq:otto-cycle-coefficients}. 
The figure plots $\beta_c^2P/\Gamma_c$ against $\eta$. 
Solid curves show the power, and dashed curves show the positive root obtained by setting Eq.~\eqref{eq:otto-quadratic-bound} to equality. 
For all three coupling ratios, the power remains positive at $\eta=\eta_C$ and over a range of $\eta>\eta_C$.

\FloatBarrier